\documentclass[12pt]{iopart}

\usepackage{subfigure}
\usepackage{rotating}

\newcommand{\msun}{M_\odot}
\newcommand{\be}{\begin{equation}}
\newcommand{\ee}{\end{equation}}
\newcommand{\bea}{\begin{eqnarray}}
\newcommand{\eea}{\end{eqnarray}}
\newcommand\flow{f_{\rm low}}
\newcommand\fisco{f_{\rm isco}}
\newcommand\fmerg{f_{\rm merg}}
\newcommand\fring{f_{\rm ring}}
\newcommand\fcut{f_{\rm cut}}
\newcommand\IMR{{\cal IMR}}
\newcommand\Iisco{{\cal I}_{\rm isco}}
\newcommand\Imerg{{\cal I}_{\rm merg}}
\newcommand\Mcrit{M_{\rm crit}}

\newcommand{\prd}{{\it Phys. Rev.} D }
\newcommand{\prl}{{\it Phys. Rev. Letter}}
\newcommand{\cqg}{\it Class. Quantum Grav. }
\newcommand{\apj}{\it Astro phys. journal}

\begin{document}

\title[]{Parameter estimation using a complete signal and inspiral templates for low mass binary black holes with Advanced LIGO sensitivity}

\author{Hee-Suk Cho}
\ead{chohs1439@pusan.ac.kr}
\address{Korea Institute of Science and Technology Information, Daejeon 305-806, Korea}

\begin{abstract}
We study the validity of inspiral  templates in gravitational wave data analysis with Advanced LIGO sensitivity
for low mass binary black holes with total masses of $M \leq 30 \msun$.
We mainly focus on the nonspinning system.
As our complete  inspiral-merger-ringdown waveform model ($\IMR$), we assume the phenomenological model, ``PhenomA",
and define our inspiral template model ($\Imerg$) by taking the  inspiral part into account from $\IMR$ up to the merger frequency ($\fmerg$).
We first calculate the {\it true} statistical uncertainties using $\IMR$ signals and $\IMR$ templates.
Next, using $\IMR$ signals and $\Imerg$ templates, we calculate fitting factors and systematic biases,
and compare the biases with the {\it true} statistical uncertainties.
We find that the valid criteria of the bank of  $\Imerg$ templates are obtained as $\Mcrit \sim 24 \msun$ for detection  (if $M>\Mcrit$, the fitting factor is smaller than $0.97$),
and $\Mcrit \sim 26 \msun$ for parameter estimation (if $M>\Mcrit$, the systematic bias is larger than the {\it true} statistical uncertainty where the signal to noise ratio is $20$), respectively.
In order to see the dependence on the cutoff frequency of the inspiral waveforms,
we define another inspiral model $\Iisco$ which is terminated at the innermost-stable-circular-orbit frequency ($\fisco<\fmerg$).
We find that the valid criteria of the bank of   $\Iisco$ templates are obtained as $\Mcrit \sim 15 \msun$ and  $\sim 17 \msun$  for detection and parameter estimation, respectively.
We investigate the statistical uncertainties for the inspiral template models considering various signal to noise ratios,
and compare those to the {\it true} statistical uncertainties.
We also consider the aligned-spinning system with fixed mass ratio ($m_1/m_2=3$) and spin ($\chi=0.5$) by employing the recent phenomenological model, ``PhenomC".
In this case, we find that the {\it true} statistical uncertainties can be much larger than those for the nonspinning system due to the mass-spin degeneracy.
For inspiral PhenomC templates truncated at $\fmerg$,  the fitting factors can be better but the biases are found to be much larger compared to those for the nonspinning system.
In particular, we find significantly asymmetric shapes of the three-dimensional overlaps including bimodal distributions. 
\end{abstract}

%=======	Title		================================	

%=======	Intro		================================	
\section{Introduction}
The next generation gravitational wave  detectors, such as Advanced LIGO~\cite{Aas15}  and Virgo,~\cite{Ace15}
are likely to allow us to observe the real signals in coming years
~\cite{Aas13b}.
Coalescing binary black holes (BBHs) are among the most promising sources 
of gravitational wave transients for the ground-based detectors~\cite{Rod15}.
A coalescing BBH system suffers three phases:
inspiral-merger-ringdown (IMR).
In the inspiral phase, the two compact objects move in quasicircular orbit
mutually approaching driven by radiation reaction.
In the merger-ringdown (MR) phases,
the system reaches the ultra-relativistic regime,
the two bodies merge to form a single excited Kerr BH
and eventually that settles down into a Kerr  BH.
The gravitational waveforms from the early inspiral phase 
can be accurately obtained by the post-Newtonian (PN)  
approximation (refer to Ref.~\cite{Buo09} for an overview on various PN approximants), and
in the ultra-relativistic regime the accurate waveforms 
can be calculated by the numerical relativity (NR) simulations~\cite{SXS}.
Of course, from the NR simulations one can extract the complete IMR waveforms (e.g. see Ref.~\cite{Szi15}).
However, since performing a long NR simulation is computationally very expensive,
most of those simulations have been done only in the last few orbits.
On the other hand, efforts to establish the analytic IMR waveform models for nonspinning BBHs
have been made by several authors~\cite{Aji07a,Aji08a,Aji08b,Pan08}
by means of the hybrid IMR waveforms, which can be obtained 
by combining the PN inspiral waveforms~\cite{Buo03,Bla04} and the numerical MR waveforms~\cite{Bar06,Buo07,Bar07,Spe07,Bru08,Han08}.
Further studies now allow us to have the
Fourier domain spinning IMR waveform models 
for aligned-spin~\cite{San10,Aji11} and precessing~\cite{Han14} BBHs.

For low mass compact binaries, 
whose components consist of a stellar mass BH and/or neutron star,
the inspiral phase is likely to have most of the signal power
and accurate inspiral waveforms can be computed by
the PN approximants.
Therefore, the PN inspiral waveforms have been 
generally used in the ground-based gravitational wave search~\cite{Aba12} and parameter estimation~\cite{Aas13a,Ber15}.
Especially, the stationary-phase approximated PN waveform model (called ``TaylorF2"~\cite{Buo09}) 
has been mainly used because the waveform can be given by an analytic function
in the Fourier domain.
However, if the MR phases are nonnegligible in the total signal power,
the complete IMR waveforms should be used as templates in searches not to lose a fraction of the signal to noise ratio (SNR).
As the binary mass increases,
the contribution level of MR phases to the SNR tends to
increase.
For detection purposes, thus, one can choose a critical mass ($\Mcrit$) considering both the computational advantage and the detection efficiency.
For the signals with masses $M<\Mcrit$ simple inspiral-only templates 
can be used to lower the computational cost although there can be a small loss of the SNR. 
While, for the signals with $M>\Mcrit$ complete IMR templates
should be used not to lower the detection efficiency below a certain value (typically $90 \%$).
Buonanno \etal~\cite{Buo09} and Brown \etal~\cite{Bro13} showed that $\Mcrit \sim 12 \msun$ for various PN inspiral template models
with IMR signals computed by the EOBNR model.
Ajith~\cite{Aji08b} also showed that $\Mcrit \sim 15 \msun$ for $3.5$PN TaylorT1 inspiral templates with phenomenological IMR signals.
The latter is consistent with one of our results, which we will see in section~\ref{sec3}.

In parameter estimation, 
inspiral waveform templates can produce systematic biases due to
the difference between two models for the IMR signals and the inspiral templates.
Farr \etal~\cite{Far09} investigated the overall trend of fitting factors and biases  using EOBNR signals with total masses up to $\sim 300~\msun$ and the TaylorF2 and the EOB inspiral template models by performing Monte Carlo simulations.
A similar work was carried out by Bose \etal~\cite{Bos10} 
using phenomenological IMR signals with masses in $m_{1,2}\in [13, 104] ~\msun$ and the 3.5 PN TaylorT1 inspiral templates.
Bose \etal~\cite{Bos10} calculated the overlap surfaces numerically to obtain the overall trend of fitting factor and biases, 
and compared the biases to those obtained by using the analytic approximation described in Ref.~\cite{Cut07} in the low mass region $m_{1,2} \in [5, 20] \msun$.
In this work, we also investigate systematic biases and fitting factors of inspiral template models for the nonspinning system.
We employ the phenomenological waveform model, ``PhenomA"~\cite{Aji08a} as our complete IMR model, and
consider inspiral templates constructed 
by taking only the inspiral parts into account from the original IMR waveforms.
However,  we focus on low mass BBHs in the range of $M \leq 30 \msun$ and access the validity of inspiral templates
for both detection and  parameter estimation in detail, giving the quantitative results.
Thus, our work can complement the previous studies and present some interesting new findings, which we will discuss in section~{\ref{sec.3.4}}.

On the other hand, 
astrophysical BHs are likely to have spins, that can significantly affect both the search and the parameter estimation in GW data analysis.
Therefore, our results for the nonspinning binary system cannot be directly applicable to spinning systems.
In order to see the impact of the spin on our results, we consider the aligned-spinning system with fixed mass ratio ($m_1/m_2 = 3$) and spin ($\chi = 0.5$) by employing the recent phenomenological model, ``PhenomC", and show a brief comparison between the nonspinning and the aligned-spinning systems.
We find that the degeneracy between the mass ratio and the spin can significantly affect the overlap distributions;
thus, the statistical uncertainties and the biases can be much larger than those for nonspinning binaries.

This paper is organized as follows.
In section~\ref{sec2}, we briefly review PhenomA model,
and  describe how to calculate statistical uncertainties from overlap surfaces  in the Bayesian parameter estimation framework.
In section~\ref{sec3}, using the IMR waveforms,
we investigate statistical uncertainties for signals with a SNR of $20$.
Next, assuming a bank of inspiral templates, we calculate fitting factors and  biases, 
and investigate the valid criteria of the inspiral template bank for the efficiencies in detection and parameter estimation.
A systematic study on the impact of the SNR on our analysis is also discussed.
Finally,  we  present some results for aligned-spinning binaries to see the spin effect.
A summary of this work is given in section~\ref{sec4} with future works.

%=======	Section 2: method	================================

\section{Phenomenological Waveforms and parameter estimation}\label{sec2}

\subsection{Phenomenological Waveforms}
In the past years, various phenomenological waveform models have been developed for nonspinning (PhenonA~\cite{Aji08a}), aligned-spinning (PhenomB~\cite{Aji11}, PhenomC~\cite{San10}) and precessing (PhenomP~\cite{Han14}) BBHs (these are implemented in the LSC Algorithm Library (LAL)~\cite{lal}).
For the nonspinning binary system we choose to use PhenomA, although that is not the latest phenomenological model.
Since PhenomA was designed to produce only the nonspinning waveforms, the speed of generating nonspinning waveforms using PhenomA is
faster than the speed of generating those aveforms using other models.
In addition, PhenomA is expressed by a piecewise function, 
the inspiral part of the IMR waveform can be explicitly defined.
Using more recent models might give more reliable results, but 
we expect that the dominant effect on our results is due to
the difference between the IMR signals and the inspiral-only templates for a given 
waveform model rather than its own reliability.
Therefore, the choice in waveform model will not significantly affect the results.

\begin{figure}[t] 
\begin{center}
\includegraphics[width=\columnwidth]{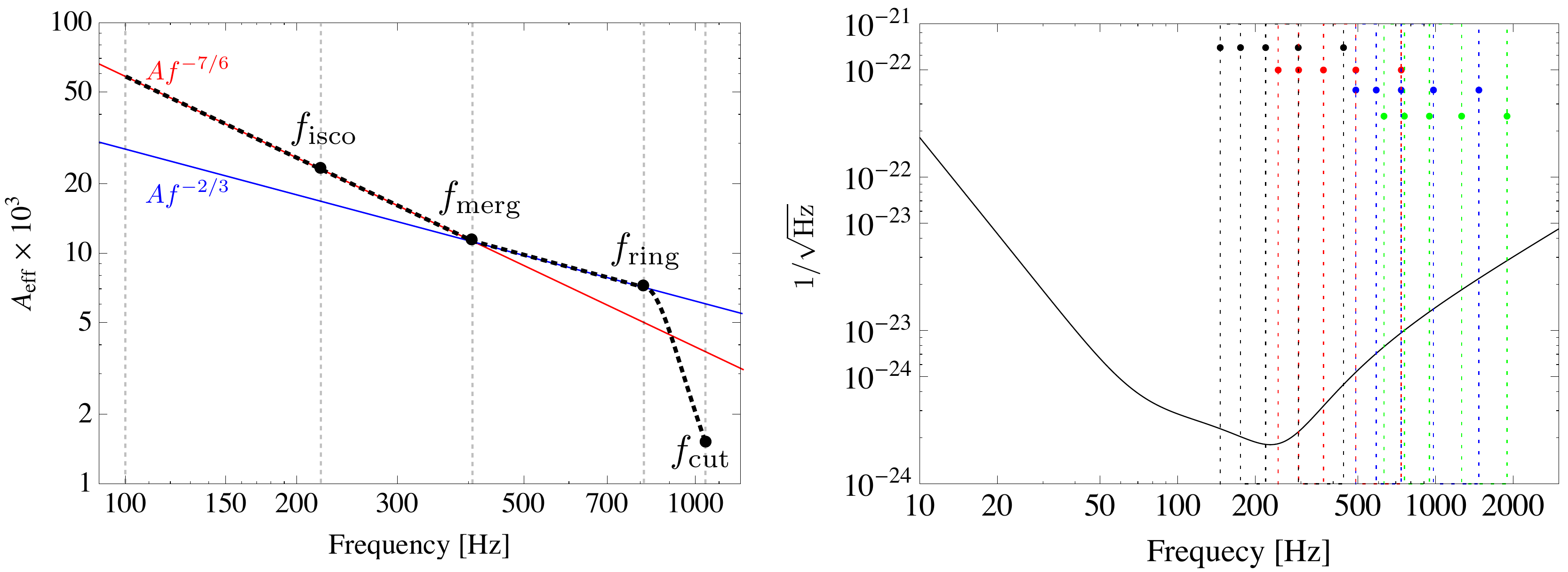}
\caption{\label{fig.spectrum} Left: Fourier domain amplitude of a (normalized) phenomenological (PhenomA) waveform starting from 100 Hz for a binary with masses $(10, 10)\msun$. Large dots indicate $\fisco$ and the phenomenological frequency parameters. Right:  The detector noise amplitude spectrum $\sqrt{S_n(f)}$ for Advanced LIGO~\cite{Aru05} and the
characteristic frequencies, $\fisco$ (black), $\fmerg$ (red), $\fring$ (blue) and $\fcut$ (green), for $M=30, 25, 20, 15$ and $10~\msun$ with $m_1/m_2=3$ from left.}
\end{center}
\end{figure}

Making use of PN-NR hybrid waveforms,
Ajith \etal~\cite{Aji08a} proposed a 
phenomenological waveform model (PhenomA) for nonspinning BBHs defined in the Fourier 
domain by,
\be
{\tilde h}_{\rm phenom}(f) = {A}_{\rm eff}(f) \, e^{\Psi_{\rm eff}(f)}.
\label{eq.hphenom}
\ee
The effective amplitude is expressed as
\bea \label{eq.Aeff}
{A_{\rm eff}} = A \fmerg^{-7/6} 
 \left\{ \begin{array}{ll}
\left(f/\fmerg\right)^{-7/6}    &\textrm{if $f < \fmerg$}   \\
\left(f/\fmerg\right)^{-2/3}    &\textrm{if $\fmerg \leq f < \fring$} \\
w \, {\cal L}(f,\fring,\bar{\sigma})  &\textrm{if $\fring \leq f < \fcut$,}\\
\end{array} \right.
\eea
where $A$ is the wave amplitude factor whose value depends on the binary masses and five extrinsic parameters determined by
the sky location and the binary orientation.
The effective phase is expressed as
\be
\Psi_{\rm eff}(f) = 2 \pi f t_c + \phi_c + \frac{1}{\eta}\,\sum_{k=0}^{7}  (x_k\,\eta^2 + y_k\,\eta + z_k) \,(\pi M f)^{(k-5)/3}\,,
\label{eq.effectivephase}
\ee
where $t_c$ and $\phi_c$ are the coalescence time and the coalescence phase,
$\eta \equiv m_1m_2/M^2$ is the symmetric mass ratio.
In equation~(\ref{eq.Aeff}),
\be
{\cal L}(f,\fring,\bar{\sigma}) \equiv \left(\frac{1}{2 \pi}\right) 
\frac{\bar{\sigma}}{(f-\fring)^2+\bar{\sigma}^2/4}\,
\ee 
is a Lorentzian function that has a  width $\bar{\sigma}$, and that is centered
around the frequency $\fring$. The normalization constant,
$w \equiv \frac{\pi \bar{\sigma}}{2} \left(\frac{f_{\rm ring}}
{f_{\rm merg}}\right)^{-2/3}$, is chosen so as to make 
${A}_{\rm eff}(f)$ continuous across the ``transition'' frequency $\fring$. 
The parameter $\fmerg$ is the frequency at which the power-law changes 
from $f^{-7/6}$ to $f^{-2/3}$. 
The phenomenological parameters $\fmerg, \fring, \bar{\sigma}$ and $\fcut$ 
are given in terms of  $M$ and  $\eta$ as
\bea \label{eq.phenomparameters}
\pi M \fmerg &=&  a_0 \, \eta^2 + b_0 \, \eta + c_0  \,, \nonumber \\
\pi M \fring &= & a_1 \, \eta^2 + b_1 \, \eta + c_1  \,, \nonumber \\
\pi M \bar{\sigma} &= & a_2 \, \eta^2 + b_2 \, \eta + c_2  \,, \nonumber \\
\pi M \fcut  &= & a_3 \, \eta^2 + b_3 \, \eta + c_3. 
\label{eq.phenomparams}
\eea
The coefficients $a_j, b_j, c_j,~j=0...3$ and $x_k,y_k,z_k,~k=0,2,3,4,6,7$ are 
tabulated in table 1 of~\cite{Aji08b}. 
The left panel of Figure~\ref{fig.spectrum} illustrates the Fourier domain amplitude spectrum and the characteristic frequencies for a binary with masses $(10,10) \msun$. In the right panel of Figure~\ref{fig.spectrum}, we show the Advanced LIGO noise curve~\cite{Aru05} which we will use in this work and compare that with the characteristic frequencies for various  binary masses.

In the past studies, the frequency cutoff of inspiral waveforms 
has been generally chosen to be the frequency at the innermost-stable-circular-orbit (isco) of the test mass orbiting a Schwarzschild black hole: 
\be \label{eq.fisco}
\pi M f_{\rm isco}=6^{-3/2}.
\ee
As in figure~\ref{fig.spectrum}, this frequency $(\fisco)$ is much smaller than the phenomenological frequency $\fmerg$.
In the range of $M<30 \msun$, the ratios $\fisco / \fmerg$ are about $0.54 - 0.64$. 
Thus,  choosing higher cutoff frequencies than $\fisco$ are more efficient to get better overlaps
with IMR signals.  
Pan \etal~\cite{Pan08} proposed the effective ringdown frequency $(f_{\rm ERD}=1.07 F_{\rm ring}$, where $F_{\rm ring}$ is the ringdown frequency of the effective-one-body waveform model) as the frequency cutoff. 
Taking the detector noise spectrum into account, Boyle \etal~\cite{Boy09} also suggested setting the frequency cutoff  
to a SNR-weighted average of $\fisco$ and $f_{\rm ERD}$.
They found that such frequency cutoffs  are more appropriate than $f_{\rm ERD}$, especially in low mass region.
In this work, we also explore the effect of the frequency cutoff by considering two inspiral template models.
We define the IMR waveform model $\IMR$, the inspiral waveform models $\Imerg$ and 
$\Iisco$, respectively as
\bea
\IMR &\equiv& {\tilde h}_{\rm phenom}(f)  \ \ \textrm{for $f\in[\flow, \fcut]$},   \nonumber\\
\Imerg &\equiv& {\tilde h}_{\rm phenom}(f)  \ \ \textrm{for $f\in[\flow, \fmerg]$}, \nonumber\\
\Iisco &\equiv& {\tilde h}_{\rm phenom}(f)  \ \ \textrm{for $f\in[\flow, \fisco]$},
\eea
where $\flow$ is the low frequency cutoff of the waveforms, that depends on the detector sensitivity.

%=======================  Parameter estimation   ============

\subsection{Parameter estimation: overlap and confidence interval} \label{sec.2.2}

In gravitational wave data analysis, a match between a signal ($\tilde{h}_s$) and a template ($\tilde{h}_t$)  is expressed by a standard inner product weighted by a detector noise power spectrum ($S_n$) as~\cite{Fin92,Dam98}
\be \label{eq.conventionaloverlap}
\langle \tilde{h}_s | \tilde{h}_t \rangle = 4 {\rm Re} \int_{\flow}^{\infty}  \frac{\tilde{h}_s(f)\tilde{h}^*_t(f)}{S_n(f)} df.
\ee
We adopt the analytic fit to the Advanced LIGO noise curve derived by Ref.~\cite{Aru05} as a form,
\begin{equation} \label{eq.ALIGOnoise}
S_n(f) = 10^{-49} \left [x^{-4.14} - 5 x^{-2} + 111 \Big(\frac{1 - x^2 + x^4/2}{1 + x^2/2}\Big)\right]\,,
\end{equation}
where $x=f/f_0$ with $f_0=215$ Hz. This noise curve is described in figure~\ref{fig.spectrum}, and we choose $\flow=10$ Hz.

In this work, we describe a single detector analysis and use the normalized waveforms,
$\hat{h}(f) \equiv \tilde{h}(f)/ \langle \tilde{h}| \tilde{h}\rangle^{1/2}$.
Then, since the phase rather than the amplitude is the main determining factor in our overlap calculations,
we do not take into account the five extrinsic parameters (i.e., the luminosity distance of the binary, two angles defining the sky position of the binary with respect to the detector, the orbital inclination and the wave polarization) in the amplitude  factor $A$. 
In addition, the inverse Fourier transform will compute the overlap for all possible coalescence times at once
and  by taking the absolute value of the complex number we can maximize the overlap over all possible coalescence phases (see \cite{All12} for more details).
In this maximization procedure, we apply a nearly continuous time shift by choosing a sufficiently small step size as in \cite{Cho14}.
The remaining physical parameters in the wave phase are two mass parameters $M$ and $\eta$ (the phenomenological parameters are also
defined by the mass parameters),
and in this work we use the chirp mass $M_c=M \eta^{3/5}$ instead of $M$ to have better performance for our approach. 
Finally, making use of the normalized signal $\hat{h}_s(\lambda_0)$, where $\lambda_0$ is the true value of the signal, and the normalized templates $\hat{h}_t(\lambda)$, where $\lambda_i=\{M_c, \eta\}$,
we calculate two-dimensional overlap surface as
\be\label{eq.maxoverlap}
P(\lambda) =  \max_{t_c,\phi_c}{\langle \tilde{h}_s(\lambda_0) | \tilde{h}_t(\lambda) \rangle   \over  \sqrt{\langle \tilde{h}_s(\lambda_0) | \tilde{h}_s(\lambda_0) \rangle \langle \tilde{h}_t(\lambda) | \tilde{h}_t(\lambda) \rangle }}.
\ee

\begin{figure}[t]
\begin{center}
\includegraphics[width=10cm]{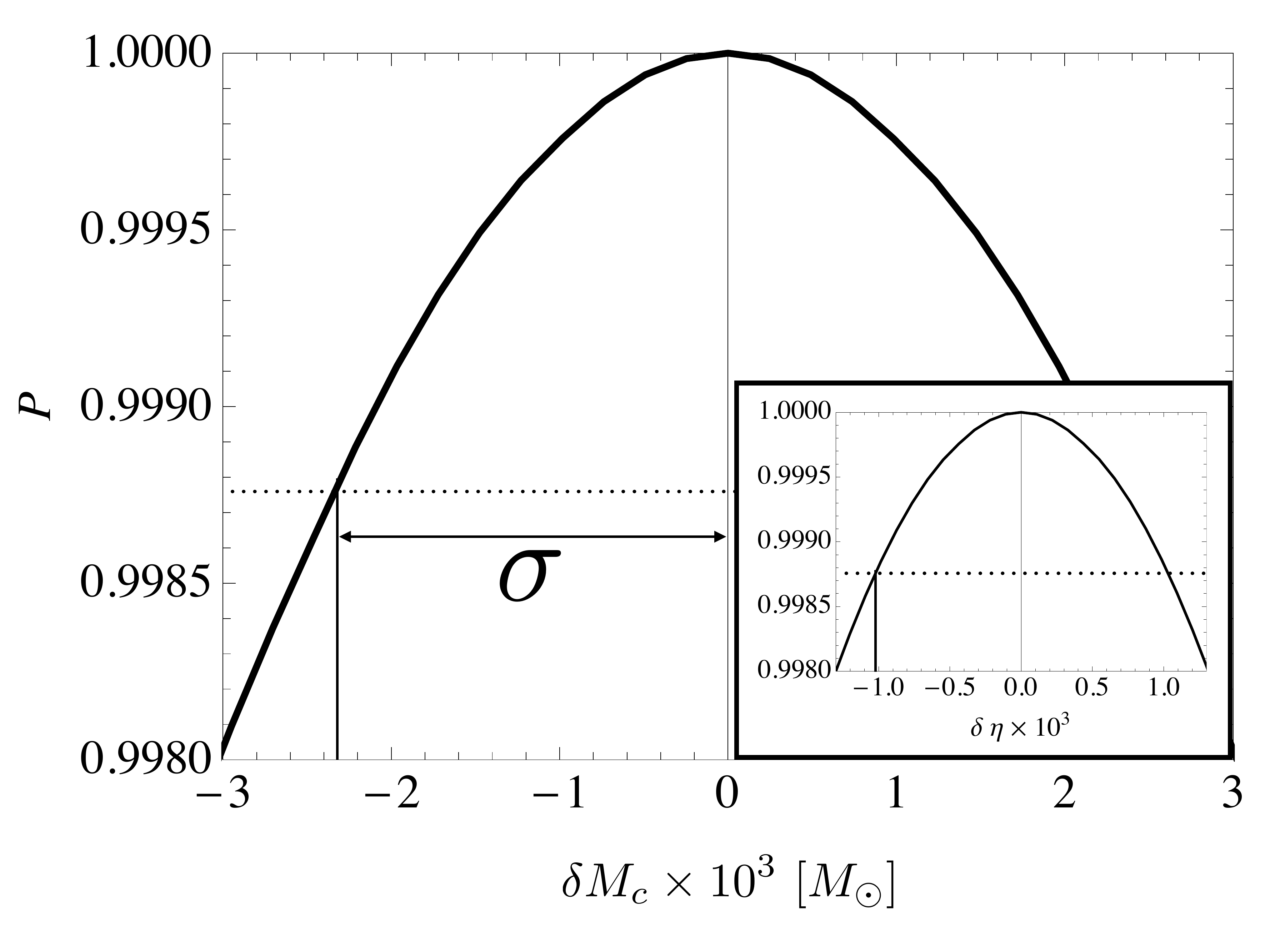}
\caption{\label{fig.CI} Schematic view showing how to calculate the (1-sigma) confidence interval ($\sigma$). 
The one-dimensional overlap distribution is calculated by marginalizing the original two-dimensional overlap surface. 
The dotted lines indicate $P=0.99876$, which corresponds to SNR=20 (see section~\ref{sec.2.2} for more details). The one-dimensional overlap distribution for $\eta$ is given in the inset.
We use a binary with masses $(15, 3) \msun$, the exact values of $\sigma_{M_c}$ and $\sigma_{\eta}$ for this binary are presented in table 1.}
\end{center}
\end{figure}

Basically,  the above overlap formalism is applied to the context of Bayesian parameter estimation.
In the high SNR limit, the likelihood ($L$) can be approximated by the overlap surface~\cite{Cho13} as\footnote{Since the likelihood is expressed by a Gaussian distribution for high SNRs, this equation implies that the high region of $P$ can be expressed by a quadratic function.}
\be \label{eq.LvsP}
\ln L(\lambda) =-\rho^2  (1-P(\lambda)),
\ee
where $\rho$ is the SNR calculated by $\rho^2=\langle \tilde{h}_s | \tilde{h}_s \rangle$.
From this relation, one might expect that the confidence region of the posterior probability density function 
is associated with a certain  region in the overlap surface.

Following the Fisher matrix (FM) formalism, 
Baird {\it et al.}~\cite{Bai13} proposed a method to calculate the confidence region from the overlap based on the iso-match contours (IM).
The connection between the confidence region and the overlap surface can be given by 
\be \label{eq.P}
P\geq1-{\chi^2_k(1-p) \over 2 \rho^2 },
\ee
where $\chi^2_k(1-p)$ is the chi-square value for which there is $1-p$ probability of obtaining that value or larger and the $k$ denotes the degree of freedom,  given by the number of parameters.
In order to calculate the confidence interval for each parameter, we will consider one-dimensional overlap distributions (i.e., $k=1$), 
those can be obtained by marginalizing the two-dimensional overlap surfaces (this marginalization can be done by projecting the two-dimensional overlap surfaces onto each parameters axis).
Since the IM method follows the FM formalism in determining confidence regions assuming flat priors,
the validity of this method relies on the Gaussianity of the likelihood~\cite{Has15}.
As in equation~(\ref{eq.LvsP}), a Gaussian likelihood corresponds to a quadratic overlap at the region given by the SNR,
and we found that all overlap surfaces were sufficiently quadratic at the region of SNR=20.
Therefore, we expect that the IM method is reliable in our approach.  

For a SNR of 20, 
the (1-sigma) confidence interval ($\sigma$) of the parameter $\lambda$
is determined by the distance between the signal $\lambda_0$ and the template $\lambda_t$, 
\be \label{eq.1d-interval}
\sigma = |\delta \lambda|\equiv |\lambda_t - \lambda_0 |,
\ee 
when the parameter value of $\lambda_t$ satisfies $P(\lambda_t)\simeq 0.99876$.
In figure~\ref{fig.CI}, we describe how to calculate $\sigma$ from the overlap  $P(\lambda)$.
As seen in this figure, the overlap distribution is almost exactly quadratic (thus symmetric) in this region.
Thus, we only consider the one-sided overlaps, where $\lambda_t \leq  \lambda_0$, in our calculations of $\sigma$.
This choice is also because the overlap cannot be calculated in the region beyond the physical boundary $\eta=0.25$.
When $\eta_0$ is very close to $0.25$, the overlap surface can be obtained only in the region $\lambda_t \leq  \lambda_0$.

When applying the IM method to the overlap surface,
one should note that a one-dimensional confidence interval ($\sigma$) and a two-dimensional confidence region
are obtained from {\it different} overlap regions.
For a given overlap surface with a SNR of 20, $\sigma$ calculated by equation (\ref{eq.1d-interval})
correspond to a one-sided width of the original two-dimensional overlap contour $P=0.99876$ ($k=1$) because the marginalised overlap distribution can be determined by projecting the two-dimensional overlap surface,
while the two-dimensional confidence region directly corresponds to the contour $P=0.99714$ ($k=2$).
Therefore, $\sigma$ is not the same as the one-sided width of the two-dimensional confidence region,
and generally $\sigma$ is smaller.

On the other hand, $\sigma$ for each parameter can be obtained directly from the original two-dimensional overlap surface without marginalizations by means of the Effective Fisher matrix method proposed by~\cite{Cho14,Cho13}. In this approach, the FM can be obtained by using an analytic function, that is calculated by fitting directly to the original high dimensional overlap surface.
O'Shaughnessy \etal~\cite{Osh14a,Osh14b} showed that the statistical uncertainties for mass parameters obtained by this method are in good agreement with the results of Bayesian Monte Carlo simulations for a nonspinning and a aligned-spin binaries~\cite{Osh14a} and a precessing binary~\cite{Osh14b}.

%=======	Section 3: result	================================
\section{Result: statistical uncertainty, fitting factor and systematic bias} \label{sec3}
In this section, we assume $\IMR$ as a complete signal model, 
and consider $\IMR$ template model to obtain {\it true} statistical uncertainties, and $\Imerg$ and $\Iisco$ template models
to calcultate fitting factors and systematic biases.
We examine the efficiencies of the inspiral template models for both the detection and the parameter estimation.
We use Advanced LIGO detector noise power spectrum defined in equation~(\ref{eq.ALIGOnoise}).
As described above, we choose a high SNR of $20$ to ensure that the Gaussian approximation holds
for IM and FM approaches in calculating statistical uncertainties.
However, we note that the fitting factor and the statistical bias are independent of SNR.

%=======	IMR-IMR		================================
\subsection{$\IMR$ templates: statistical uncertainties}

\begin{figure}[t]
\begin{center}
\includegraphics[width=7cm]{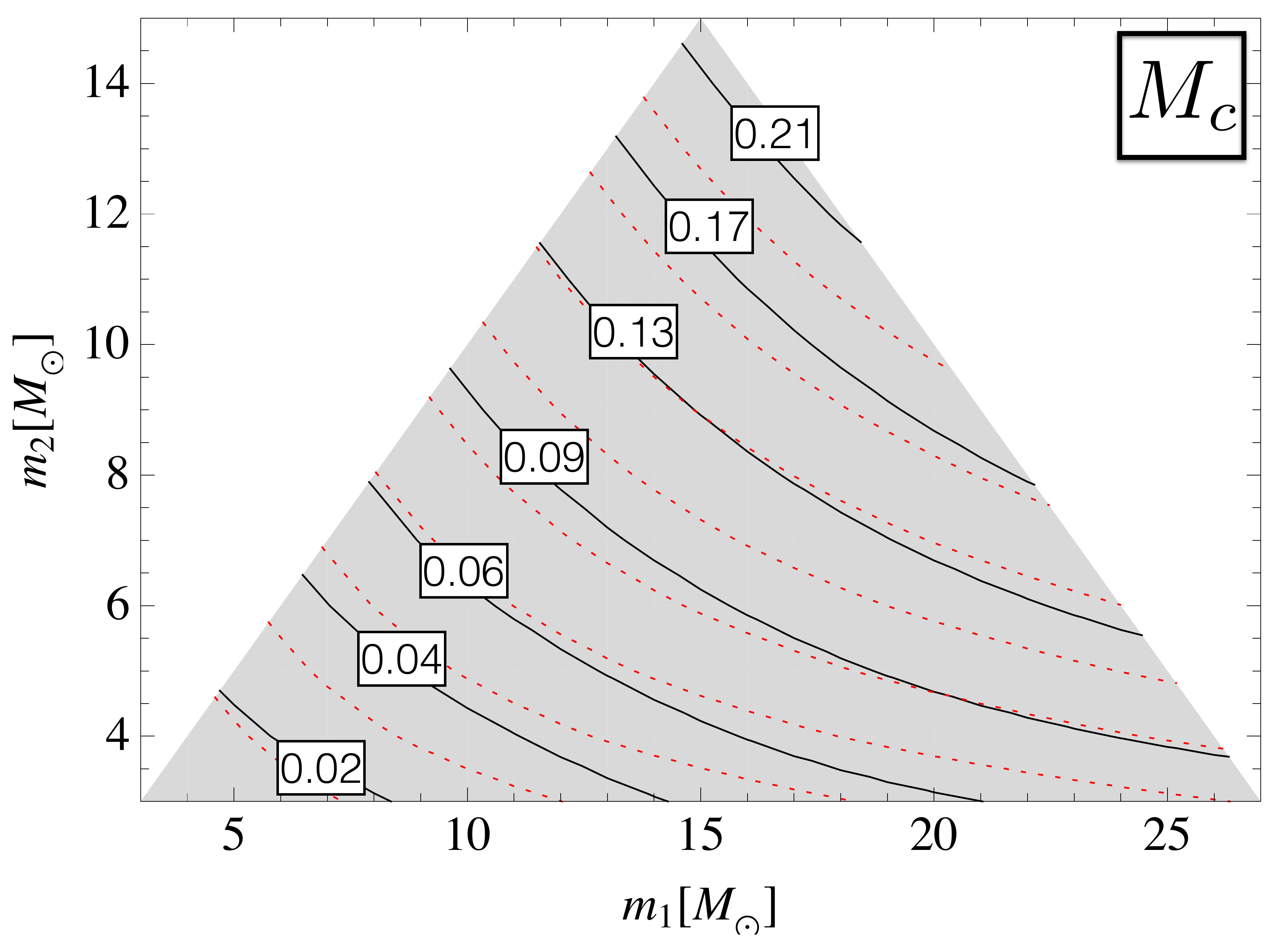}
\includegraphics[width=7cm]{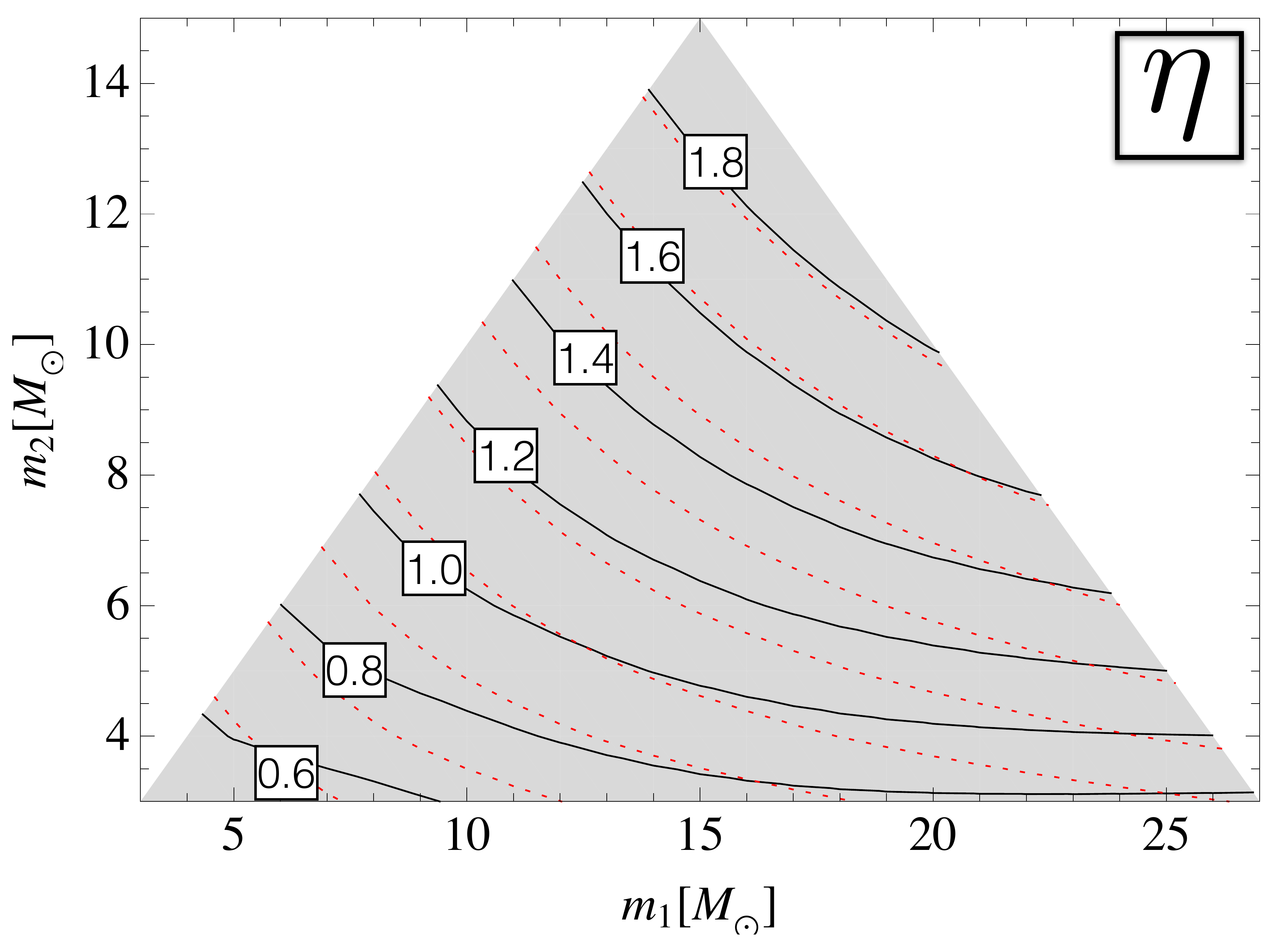}
\caption{\label{fig.error} Statistical uncertainties in percentage ($100 \times \sigma_{\lambda} / \lambda$) for nonspinning BBHs with a SNR of $20$. 
Red dotted lines indicate the constant chirp masses, $M_c=\{4, 5, ..., 12\} \msun$ from bottom left.}
\end{center}
\end{figure}

In order to obtain the  statistical uncertainty in parameter estimation,
we first consider $\IMR$ templates. 
We calculate overlap surfaces for nonspinning BBHs with 
masses $m_{1,2} \geq 3 \msun$ and $M \leq 30 \msun$, from which 
the confidence intervals are obtained by using the IM method described in figure~\ref{fig.CI}.
The percentage uncertainties ($100 \times \sigma_{\lambda} / \lambda$) are summarised in figure~\ref{fig.error}.
In the left panel, 
the uncertainty contours are overall aligned with the constant chirp mass curves (red dotted lines)
and slightly misaligned in highly asymmetric mass region.
The percentage uncertainties for $M_c$ range broadly from $\sim 0.0067 \%$ to $\sim 0.22 \%$
depending on the chip mass of the signal.
In the right panel, the pattern of contours is overall similar to the case for $M_c$,
but the contours are more misaligned to the chirp curves  in highly asymmetric mass region.
The percentage uncertainties for $\eta$ seem to increase linearly depending on the chirp mass from $\sim 0.43 \%$ to $\sim 2.0 \%$.
Overall, the accuracy of parameter estimation for $\sigma_{M_c} / M_c$ is better than that for $\sigma_{\eta} / \eta$
roughly by $1-2$ orders of magnitude (c.f. figure 6 in Ref.~\cite{Aji09}).

In the high SNR limit, on the other hand, 
the Fisher matrix (FM)  can be used to 
estimate the statistical uncertainty (refer to Ref.~\cite{Val08} and references therein for more details).
The FM is defined by
\be \label{eq.analyticfisher}
\Gamma_{ij}=\bigg\langle {\partial \tilde{h} \over \partial \lambda_i} \bigg | {\partial \tilde{h} \over \partial \lambda_j} \bigg \rangle\bigg|_{\lambda=\lambda_0},
\ee
where $\lambda_0$ is the true value of the signal.
Since Fourier-domain waveform models can be
expressed by analytic functions of the parameters,
the derivatives in this equation can be obtained analytically, and
only one overlap computation has to be performed numerically to obtain the FM.
Thus, the FM method is computationally much cheaper than the IM method.
In addition, since the FM formalism is also constructed under the assumption of the  Gaussian likelihood,
we anticipate that that gives the results similar to our statistical uncertainties calculated by the IM method.
Taking the phenomenological IMR waveform model 
into account, we examined the accuracy of the analytic FM
by comparing the FM estimations to our results, and
found that both results were
in very good agreement 
within $\sim 2 \%$ differences for all binaries considered in this work. 
In table~\ref{tab.fishercomparison},  we present the comparison results for several binaries.

\begin{table}[t]
\caption{\label{tab.fishercomparison}{Statistical uncertainties computed by using the analytic FM and the numerical IM method for nonspinning BBHs with a SNR of $20$ using the phenomenological IMR waveform model.}}
\begin{indented}
\item[]\begin{tabular}{c  ccccccccccccccccccc  }
\br
$m_1, m_2$  &\multicolumn{2}{c}{ $3\msun, 3\msun$ }& &\multicolumn{2}{c}{ $15\msun, 3\msun$}&&\multicolumn{2}{c}{ $27\msun, 3\msun$ } &&\multicolumn{2}{c}{ $15\msun, 15\msun$} \\
\cline{1-3}
\cline{5-6}
\cline{8-9}
\cline{11-12}
  Method          & IM    &FM   && IM       &FM  &               & IM       &FM   && IM       &FM          \\
\mr
 $\sigma_{M_c} \times 10^{4} [\msun]$	 &1.75  &1.72      &&23.3  &23.0&     &52.3  &51.9&&289  &287  \\
 $\sigma_{\eta}  \times 10^{4}$	  &10.7 &10.7      &&10.2  &10.2&&     6.91  &6.98&&49.1  &50.1\\
 \br
 \end{tabular}
 \end{indented}
\end{table}

%=================   IMR-Imerg	====================
\subsection{$\Imerg$ templates: fitting factors and systematic biases} \label{sec.3.2}

\begin{figure}[t]
\begin{center}
\includegraphics[width=10cm]{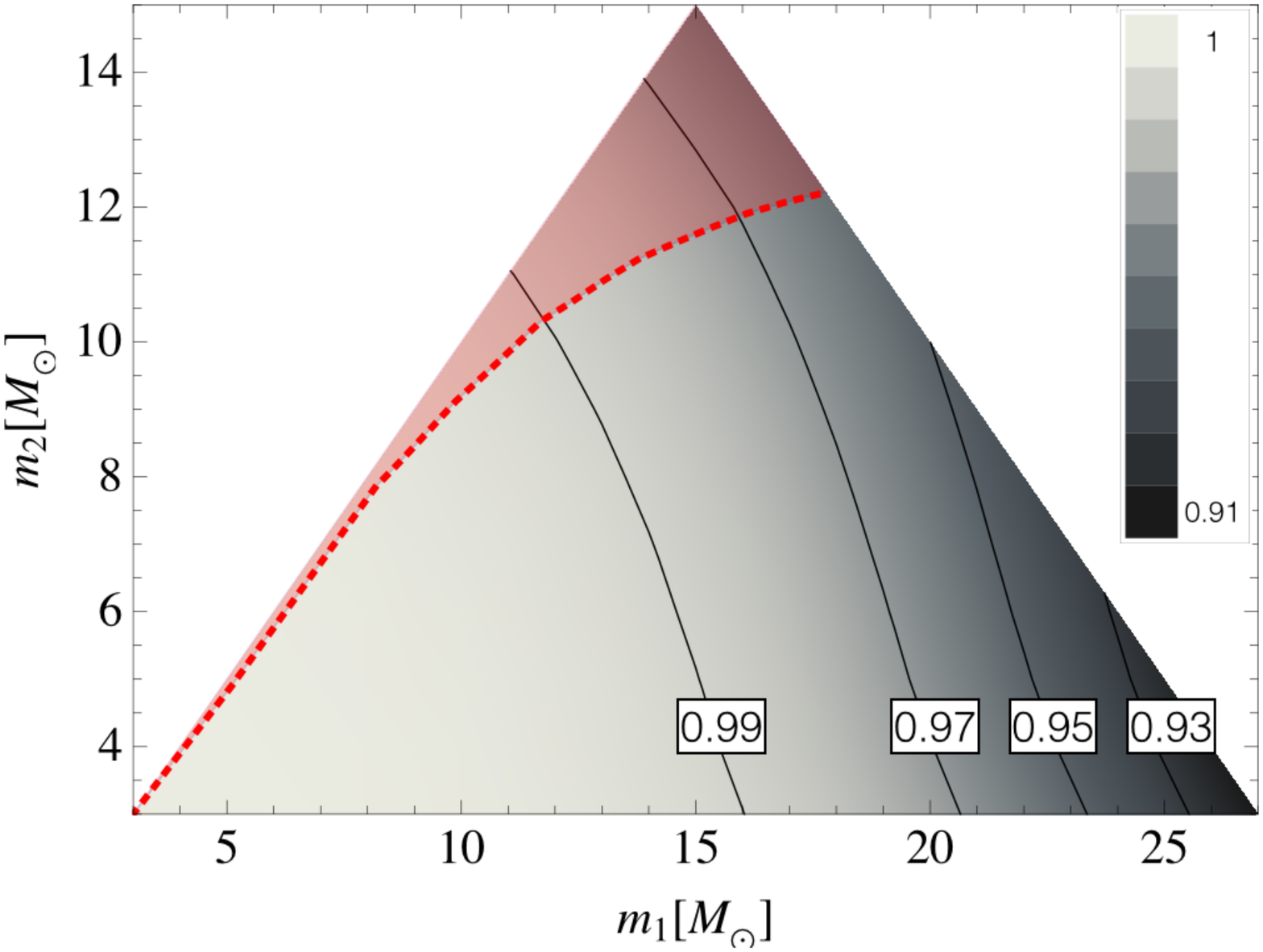}
\caption{\label{fig.ff-merg}Fitting factors for $\Imerg$ templates for nonspinning BBHs. 
The red dotted line denotes $\eta_{\rm crit}$ (see section~\ref{sec.3.2} for more details).}
\end{center}
\end{figure}

Next, we consider $\Imerg$  as a template model
to investigate fitting factors and systematic biases.
When measuring the match between two different waveform models,  
the fitting factor (FF) or equivalently the mismatch ($1-{\rm FF}$),
is widely used~\cite{Buo09,Aji08b,Bro13,Apo95}:
\be
{\rm FF} = \max_{t_c,\phi_c,\lambda_i}{\langle \tilde{h}_s(\lambda_0) | \tilde{h}_t(\lambda) \rangle   \over  \sqrt{\langle \tilde{h}_s(\lambda_0) | \tilde{h}_s(\lambda_0) \rangle \langle \tilde{h}_t(\lambda) | \tilde{h}_t(\lambda) \rangle }}.
\label{FF}
\ee 
FF is the normalized overlap between a signal waveform $h_s(\lambda_0)$ and a set of template
waveforms $h_t(t_c, \phi_c,\lambda_i)$ maximized over $t_c, \phi_c$ and other parameters $\lambda_i$. 
Thus, in this work, FF corresponds to the maximum value
of the  overlaps,
\be
{\rm FF} \equiv \max_{M_c, \eta} P(M_c, \eta).
\ee
In gravitational wave data analysis, FF is used to evaluate
the detection efficiency.
The gravitational wave searches use a bank of template waveforms
constructed for the corresponding mass range of the systems~\cite{Apo95,Sat91,Bal96,Owe96}.
Typically, a template bank requires that the  mismatch between the templates and
the signal does  not  exceed $3\%$~\cite{Aba12,Aas13} including the effect of the discreteness of the template spacing.
In this work, we use sufficiently dense spacings in the $(M_c$-$\eta)$ plane in order to avoid this  discreteness effect\footnote{To obtain FF for one signal, for example, we repeat a grid search near the signal varying the search area and the template spacings until we can roughly estimate the size of the overlap contour $\hat{P}=0.995$, where $\hat{P}$ is an overlap weighted by the maximum overlap value in that contour, and finally we find FF by performing a  $51 \times 51$ grid search in the region $\hat{P}>0.995$~\cite{Cho15}.}.
For the two different waveform models, the SNR can be defined by
\be
\rho=\langle \tilde{h}_s | \tilde{h}_t \rangle^{1/2}=\langle \tilde{h}_s | \tilde{h}_s \rangle^{1/2}{\rm FF}.
\ee
The detection rate is proportional to the cube of the SNR (thus to the cube of FF).
Therefore, a ${\rm FF}=0.97$  corresponds to a loss of detection rates of $\sim 10\%$. 
In figure~\ref{fig.ff-merg}, we show FFs for nonspinning BBHs for $\Imerg$ templates.
We find that
the condition ${\rm FF}>0.97$ holds for the binaries in the region $M<24 \msun$,
thus we have $\Mcrit\sim 24 \msun$ for the detection efficiency. 
On the other hand, we see that FF depends on the total mass overall.
This is because the contribution level of the MR phases to the complete IMR waveform
tends to increase as the mass increases.

\begin{figure}[t]
\begin{center}
\includegraphics[width=7cm]{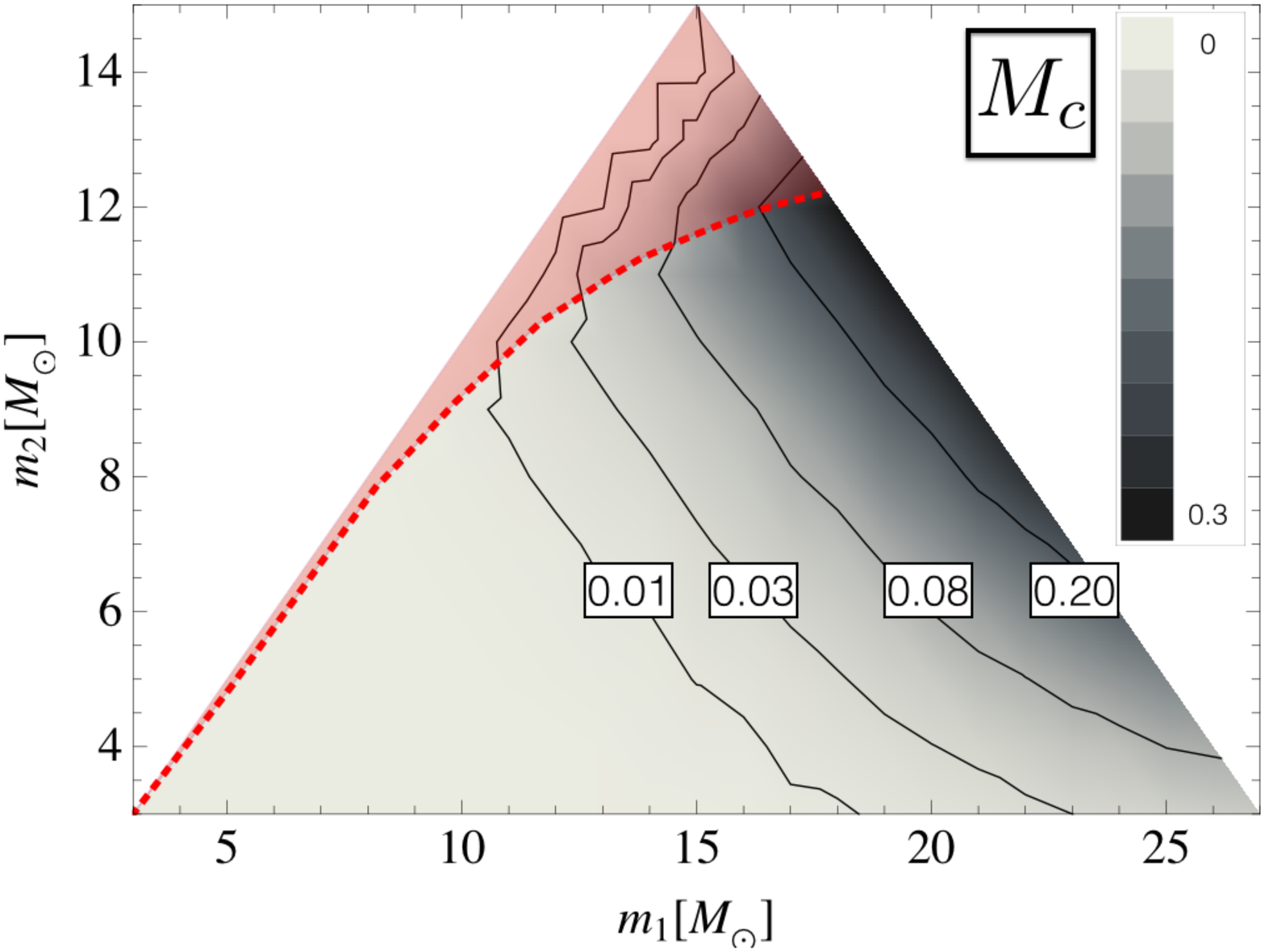}
\includegraphics[width=7cm]{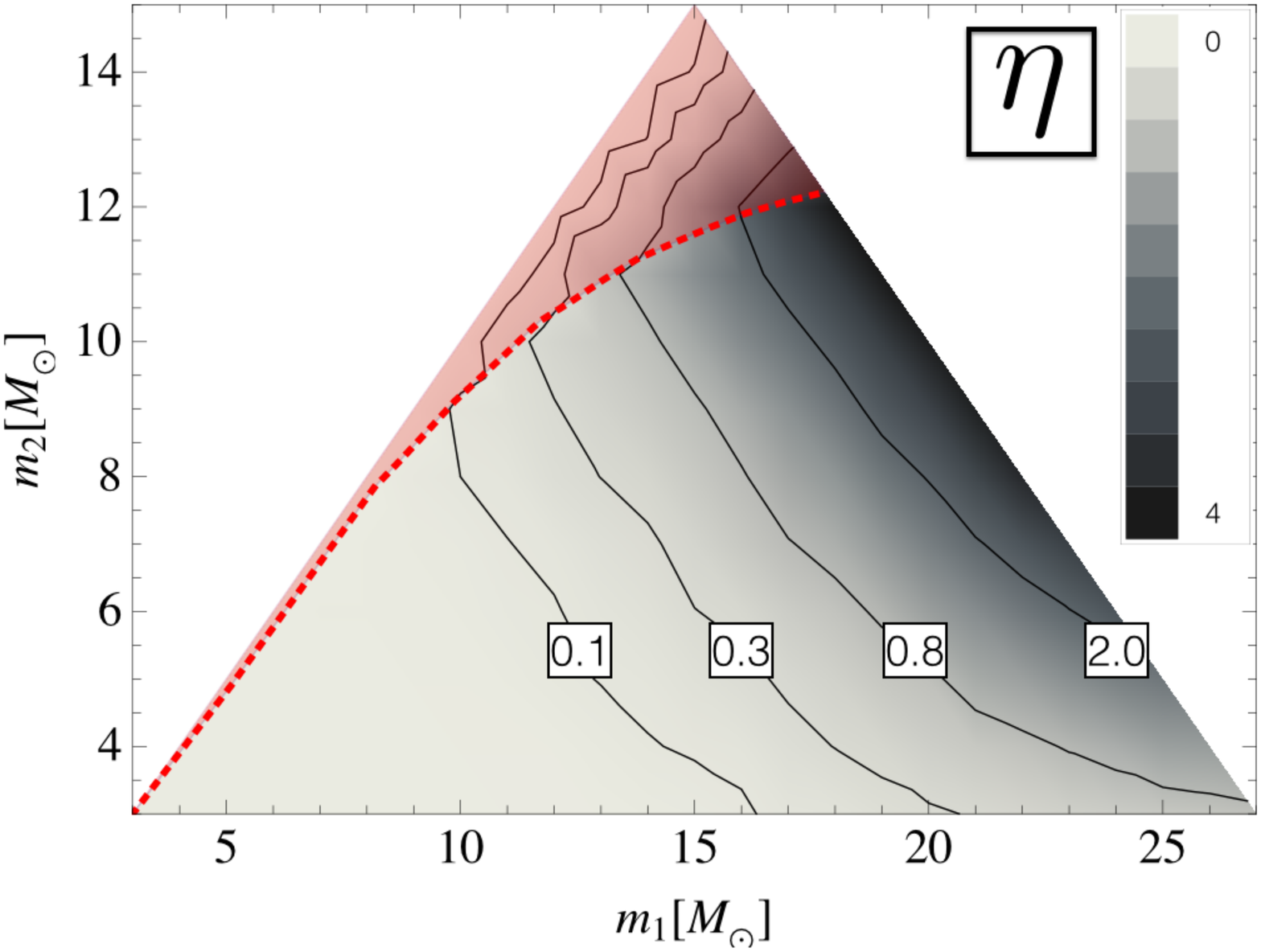}
\caption{\label{fig.bias-merg}
Systematic biases in percentage ($100 \times b_{\lambda} / \lambda$) for $\Imerg$ templates for nonspinning BBHs. The red dotted line denotes $\eta_{\rm crit}$}
\end{center}
\end{figure}

Once FF is calculated by the overlap surface as
${\rm FF}=P(M_c^{\rm rec}, \eta^{\rm rec})$,
we define the bias of the parameter $\lambda$ by the distance from the true value ($\lambda_0$) to the recovered value ($\lambda^{\rm rec}$),
\be\label{eq.bias}
b_{\lambda} =   \lambda^{\rm rec} - \lambda_0.
\ee
The percentage biases ($100 \times b_{\lambda} / \lambda$) are summarised in figure~\ref{fig.bias-merg}.
We find that the bias also depends on the total mass overall,
those can increase up to $\sim 0.3 \%$ and $\sim 4 \%$ for $M_c$ and $\eta$, respectively.
On the other hand, 
the recovered values ($\lambda^{\rm rec}$) are positive for all true values ($\lambda_0$), i.e., $b_{\lambda}>0$.
So, for a given chirp mass, $\eta^{\rm rec}$ increases with increasing $\eta_0$, and
when $\eta^{\rm rec}$ is equal to the physical boundary $0.25$, $\eta_0$ can be equal to some critical value ($\eta_{\rm crit}$),
however $\eta^{\rm rec}$ cannot exceed $0.25$ although $\eta_0$ increases over $\eta_{\rm crit}$.
Thus, in the range of $\eta_{\rm crit} \leq \eta_0 \leq 0.25$, we always have $\eta^{\rm rec}=0.25$.
In addition, since $\eta^{\rm rec}$ is fixed at $0.25$ in this range, as $\eta_0$ approaches $0.25$, the bias ($b_\eta =\eta^{\rm rec}-\eta_0=0.25-\eta_0$) approaches $0$ (where $b_{M_c}$ also approaches $0$).
These configurations are well described by the contours in the red shaded regions in this figure.
Although a value beyond the physical boundary 
implies complex-valued masses, the PN waveforms are well
behaved for $0<\eta<1.0$~\cite{Boy09}.
For detection purposes,
Boyle \etal~\cite{Boy09} showed that
allowing such unphysical values, FFs can be significantly improved  
for the binaries above 30 $\msun$.
However, since unphysical values are not allowed in parameter estimation,
we only take into account physical values for the parameter $\eta$\footnote{In real Monte Carlo simulations, the range of $\eta$ can be incorporated as a prior,
and the posterior distribution can be affected by the prior. Thus, when the true value of $\eta$ is close to 0.25, the statistical uncertainties tend to be reduced compared to the FM estimations (e.g. see figure 2 of Ref.~\cite{Cok08}).}.

\begin{figure}[t]
\begin{center}
\includegraphics[width=7cm]{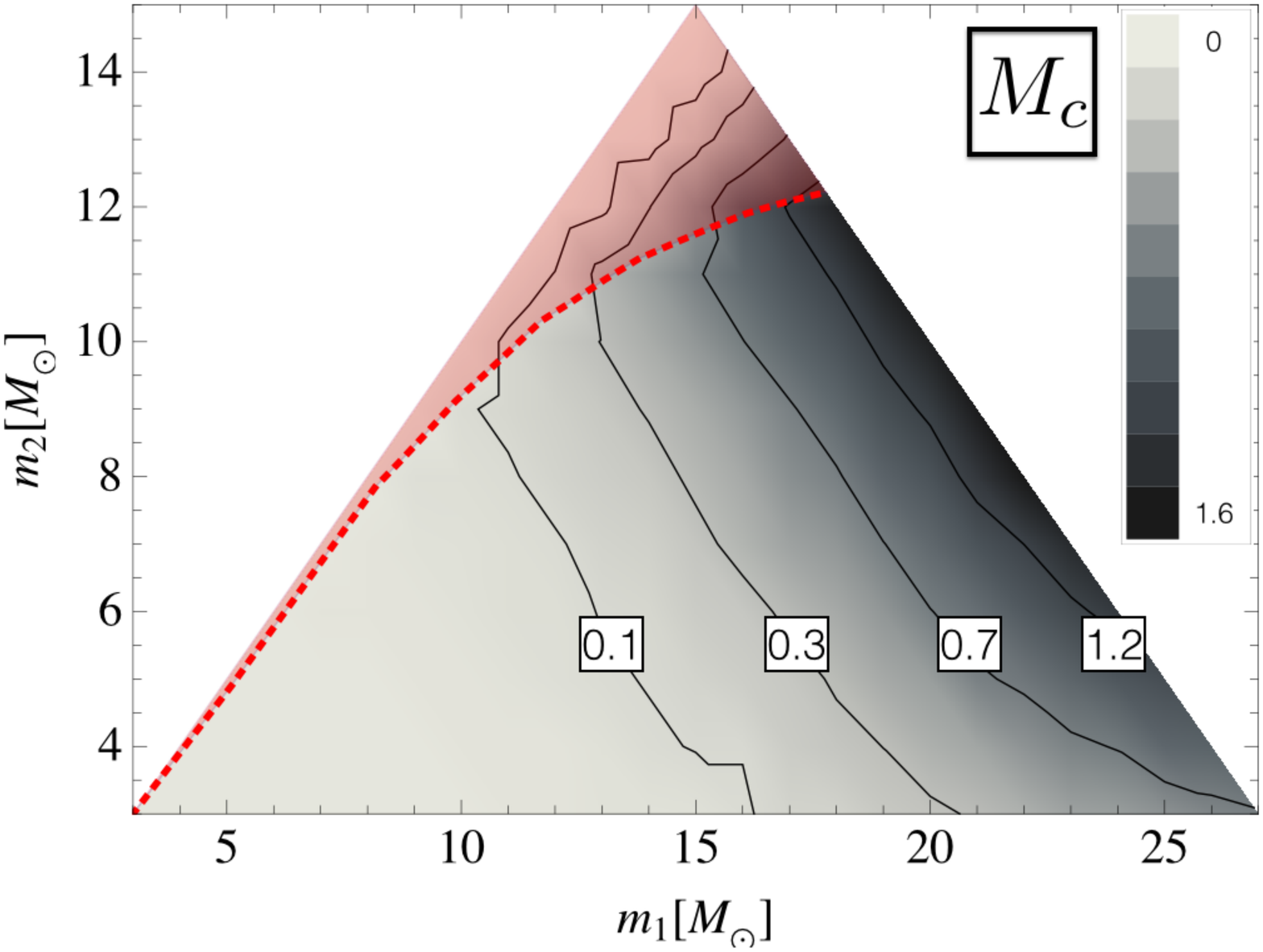}
\includegraphics[width=7cm]{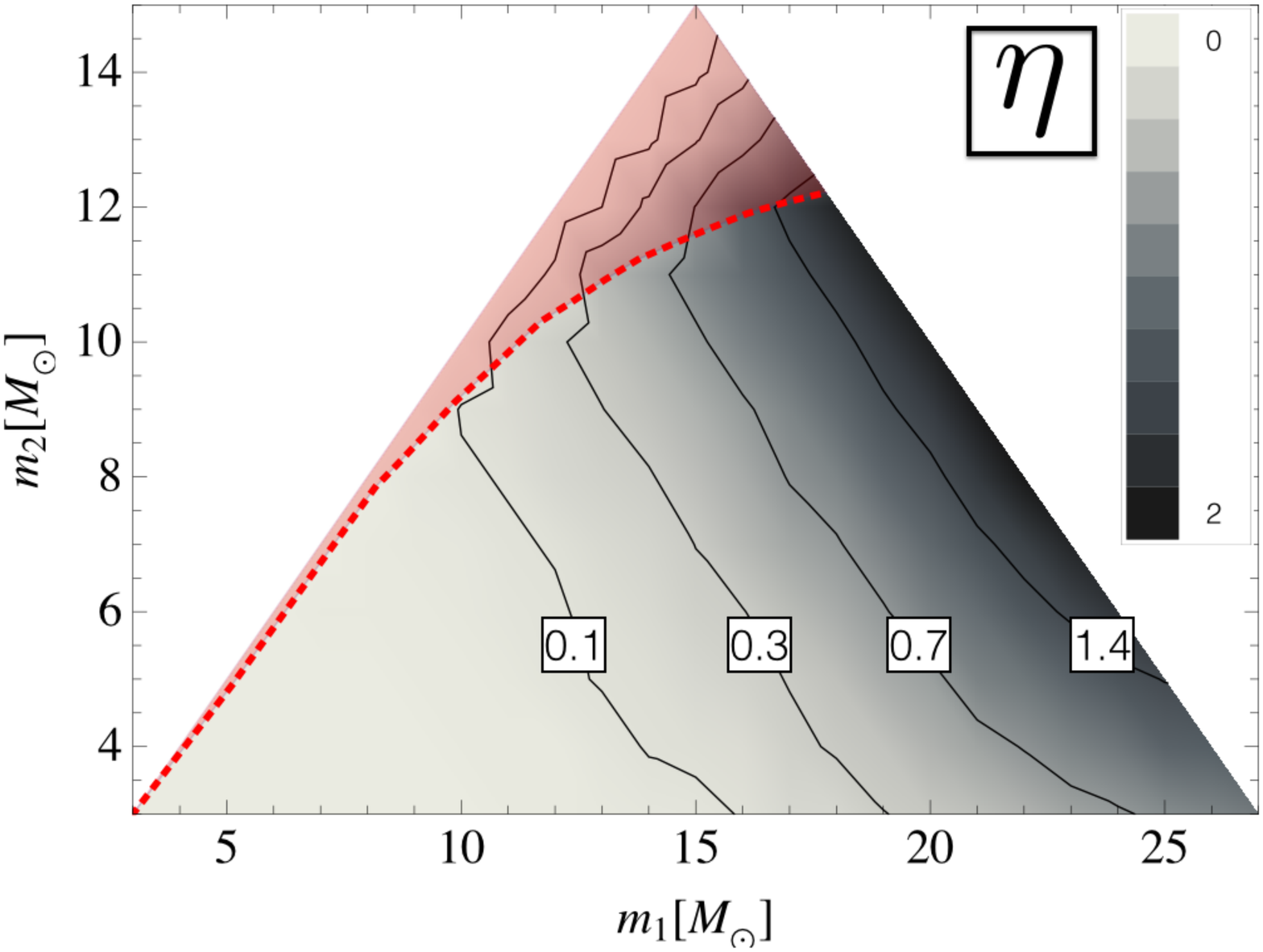}
\caption{\label{fig.frac-bias-merg}
Fractional biases ($b/\sigma$) for $\Imerg$ templates for nonspinning BBHs, where $\sigma$ is the {\it true} statistical uncertainty given in figure~\ref{fig.error}. 
The red dotted line denotes $\eta_{\rm crit}$.
}
\end{center}
\end{figure}

We have shown that both the statistical uncertainty and the systematic bias increase with increasing $M$.
However, as predicted by a simple analytic approach in~\cite{Man14},
the bias increases more rapidly than the statistical uncertainty.
In parameter estimation, a more appropriate quantity can be
the ratio of the systematic bias ($b$) and the statistical uncertainty ($\sigma$).
In figure~\ref{fig.frac-bias-merg}, we present the fractional biases ($b/\sigma$),
where $\sigma$ is the {\it true} statistical uncertainty obtained by using $\IMR$ templates as in figure~\ref{fig.error}.
We find that the fractional bias tends to exceed unity 
if the total mass is larger than $\sim 26 \msun$.
This means,
the systematic bias (produced by a simplification of the template waveforms
by taking only the  inspiral phase into account from the complete IMR phases)
becomes larger than the statistical uncertainty (calculated by using the complete IMR template waveforms for a SNR of 20).
Thus, the critical mass for $\Imerg$ templates for the  parameter estimation efficiency
is obtained as $\Mcrit\sim  26 \msun$.

%=================   IMR-Iisco	====================

\subsection{$\Iisco$ template: fitting factors and systematic biases}

\begin{figure}[t]
\begin{center}
\includegraphics[width=10cm]{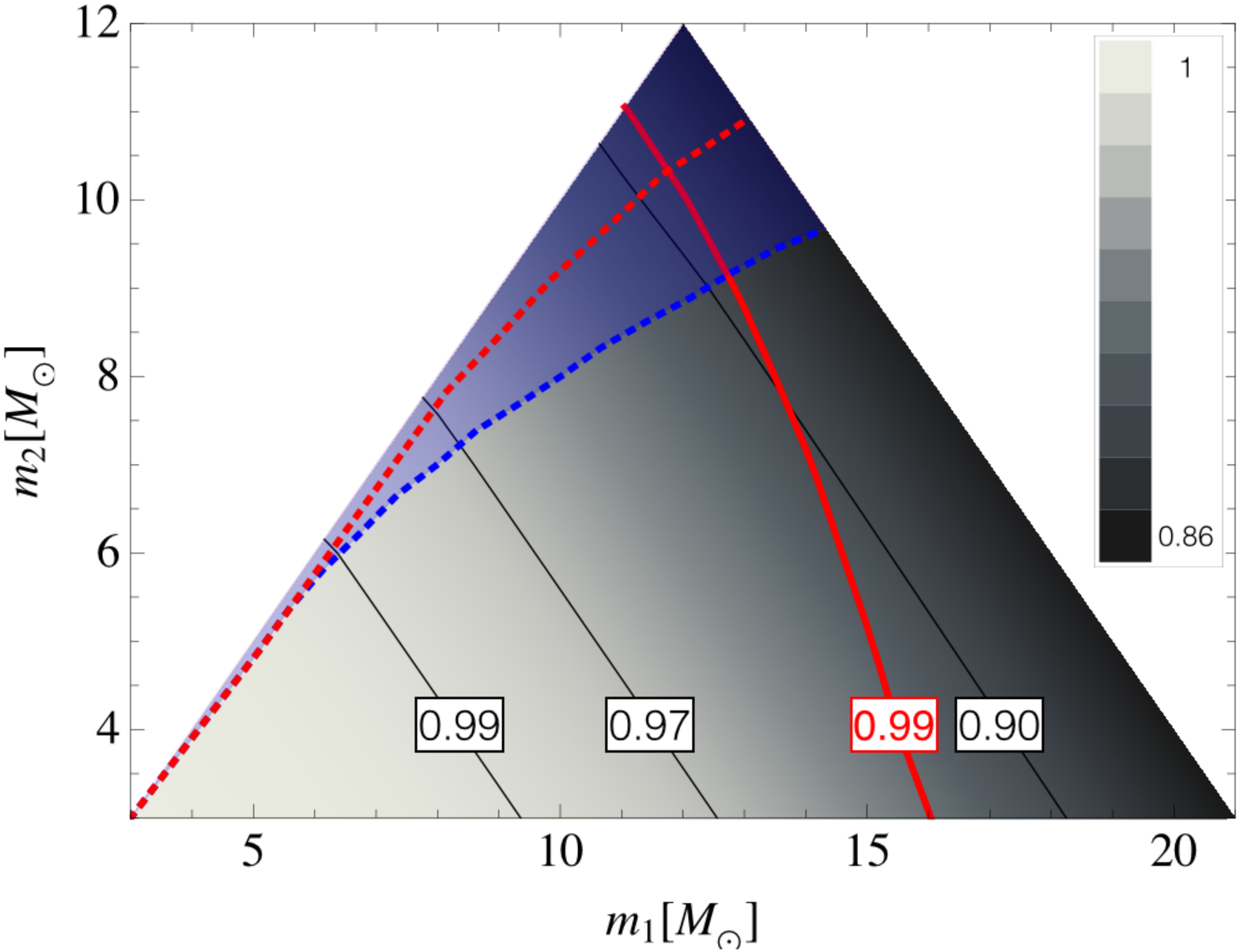}
\caption{\label{fig.ff-isco}Fitting factors for $\Iisco$ templates  for nonspinning BBHs.
The blue dotted line denotes $\eta_{\rm crit}$.
For comparison, we include $\eta_{\rm crit}$ (red dotted line) and a FF contour (red line) for $\Imerg$ templates taken from Figure~\ref{fig.ff-merg}.}
\end{center}
\end{figure}

We also take into account  $\Iisco$ as a template model,
and calculate FFs and biases.
In this model, we only consider the binaries in the range of $M\leq 24 \msun$.
FFs for this template model are given in figure~\ref{fig.ff-isco}.
We find that the dependence of FF on the total mass becomes stronger than the case for $\Imerg$, where
the contours are almost exactly parallel with the line of constant $M$.
This is because while $\fmerg$ is a function of both $M$  and $\eta$ as in equation~(\ref{eq.phenomparameters}), $\fisco$
depends only on $M$ as in equation~(\ref{eq.fisco}).
Since $\Iisco$ waveform ends the inspiral phase much more quickly than $\Imerg$ waveform,
we also find that
FFs are significantly reduced compared to those for $\Imerg$ (red line).
The valid criterion of the template bank to satisfy ${\rm FF} > 0.97$
is obtained in the range of $M<15 \msun$.
For  $\Iisco$ templates, therefore, we have $\Mcrit \sim15 \msun$ for the detection efficiency. 
Ajith~\cite{Aji08b} also found $\Mcrit \sim15 \msun$ for $3.5$PN TaylorT1 inspiral templates with $\IMR$ signals.
This consistency can be explained by that  the inspiral part of the phenomenological model has been modeled after  the $3.5$PN TaylorT1 approximant.

\begin{figure}[t]
\begin{center}
\includegraphics[width=7cm]{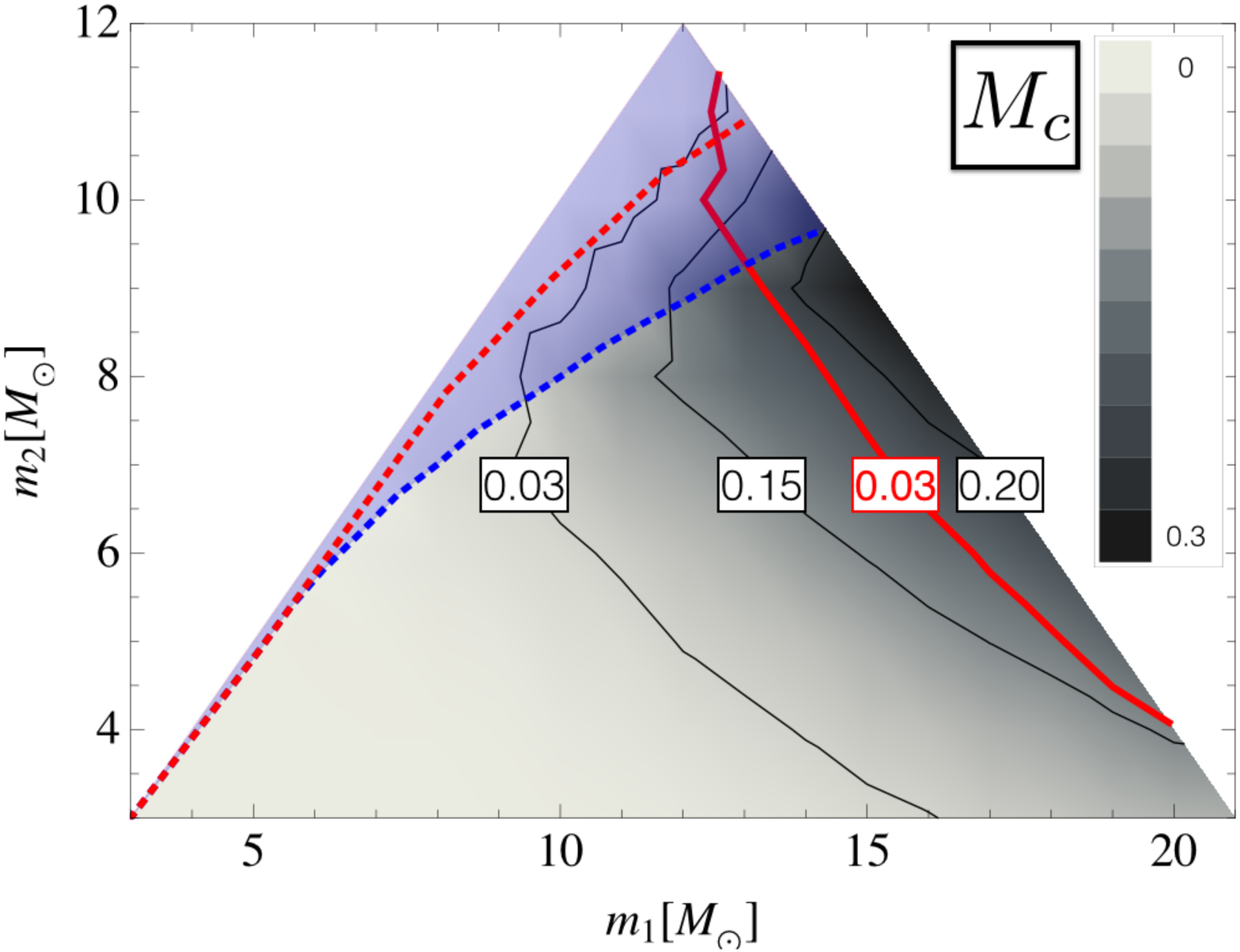}
\includegraphics[width=7cm]{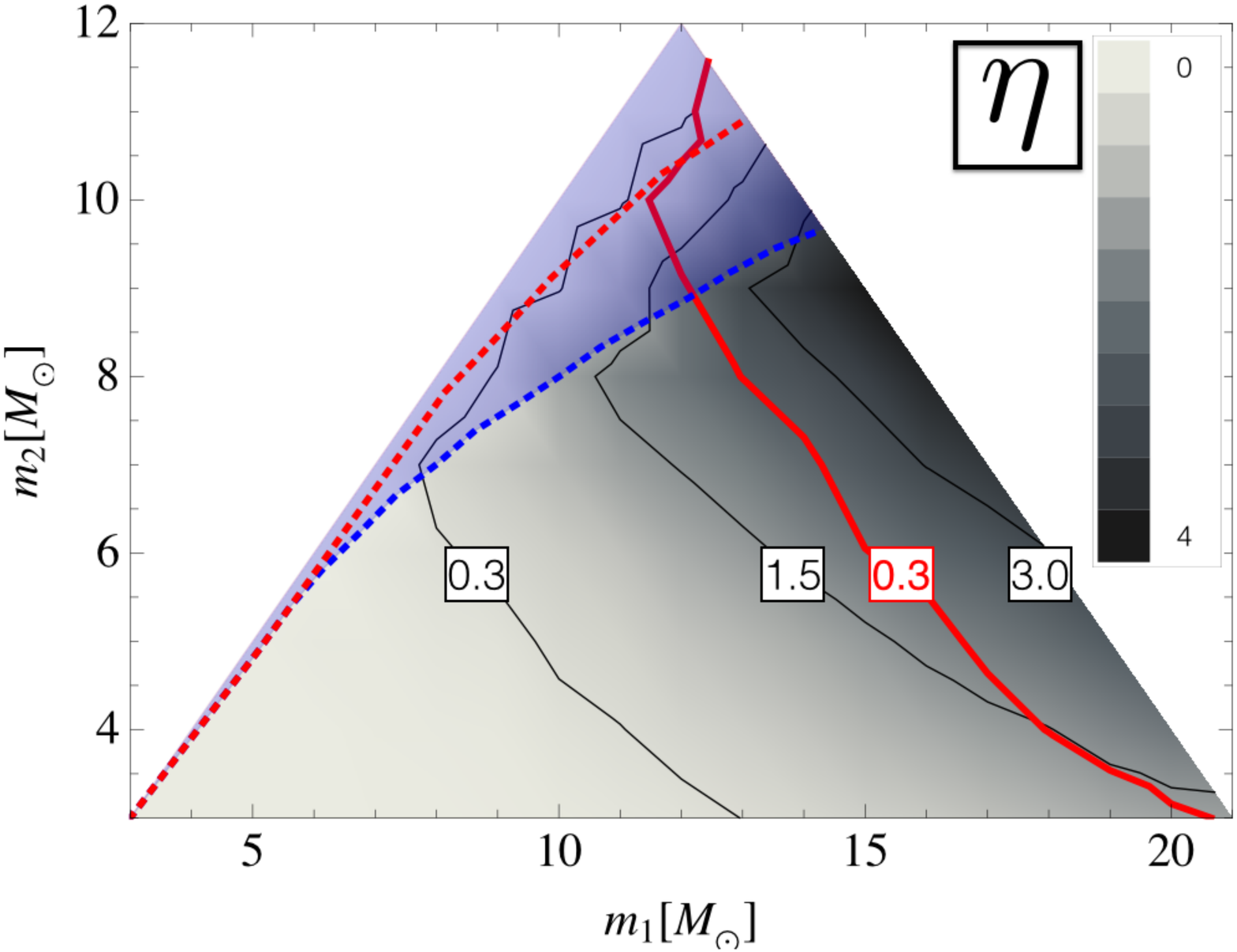}
\caption{\label{fig.bias-isco} Systematic biases in percentage ($100 \times b_{\lambda} / \lambda$) for $\Iisco$ templates for nonspinning BBHs.
The blue dotted line denotes $\eta_{\rm crit}$.
For comparison, we include $\eta_{\rm crit}$ (red dotted line) and a bias contour (red line) for $\Imerg$ templates taken from Figure~\ref{fig.bias-merg}.}
\end{center}
\end{figure}

The percentage biases ($100 \times b_{\lambda} / \lambda$) for $\Iisco$ templates
are summarised in figure~\ref{fig.bias-isco}.
We find that the biases are several times larger than those for $\Imerg$ templates,
thus the curve of $\eta_{\rm crit}$ (blue dotted line) is placed below that for $\Imerg$ templates (red dotted line).
A similar work was carried out by Bose \etal~\cite{Bos10} using $\IMR$ signals and $3.5$PN TaylorT1 templates,
employing the same numerical formalism in calculating the biases as in this work.
They took into account $M$ and $\eta$ as the mass parameters, and found an increasingly negative bias in $M$.
This is explained by the fact that
the templates that give the best fit (i.e., FF) tend to have a smaller $M$, 
which tends to increase a template's duration,
thereby compensating somewhat its lack of the MR phases~\cite{Bos10}.
While, we use $M_c$ and $\eta$ parameters, and find that the biases for both parameters are increasingly positive.  
The reason is the same, the binary mass depends on both $M_c$ and $\eta$  ($M=M_c  \eta^{-3/5}$),
and we found that a pair of the recovered values ($M_c^{\rm rec} >M_{c0}, \eta^{\rm rec}>\eta_0$) always gives a smaller $M$ than
the total mass given by the true values ($M_{c0}, \eta_0$).
For example, a binary with masses $(10, 5) \msun$ gives $\{M_0, M_{c0}, \eta_0 \}=\{15\msun, 6.0836\msun, 0.2222\}$ and
$\{M^{\rm rec}, M_c^{\rm rec}, \eta^{\rm rec} \}=\{14.9685\msun, 6.0845\msun, 0.2268\}$.  For a binary with $(15, 8) \msun$, we have $\{M_0, M_{c0}, \eta_0 \}=\{23\msun, 9.4442\msun, 0.2268\}$ and
$\{M^{\rm rec}, M_c^{\rm rec}, \eta^{\rm rec} \}=\{22.6119\msun, 9.4620\msun, 0.2341\}$.

\begin{figure}[t]
\begin{center}
\includegraphics[width=7cm]{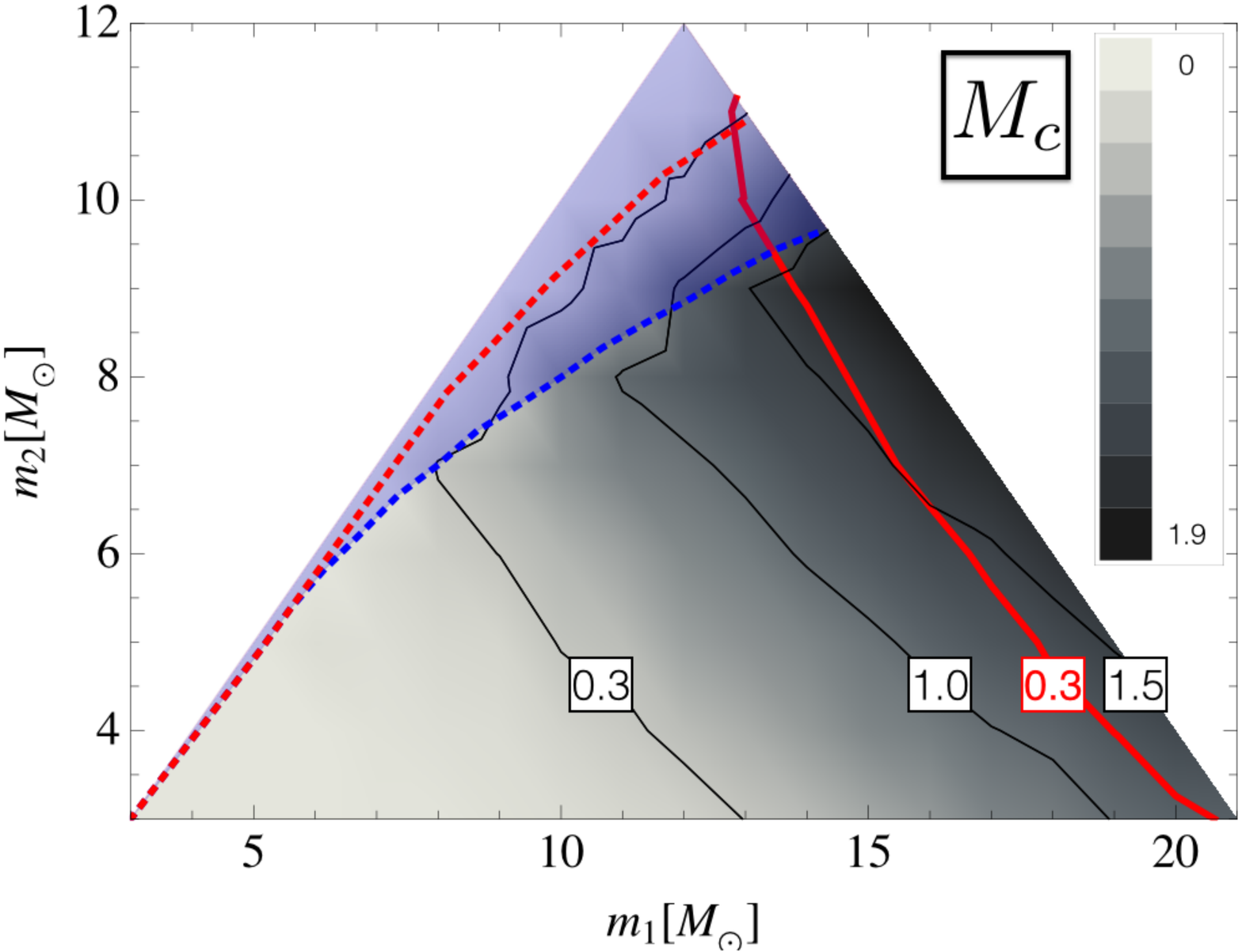}
\includegraphics[width=7cm]{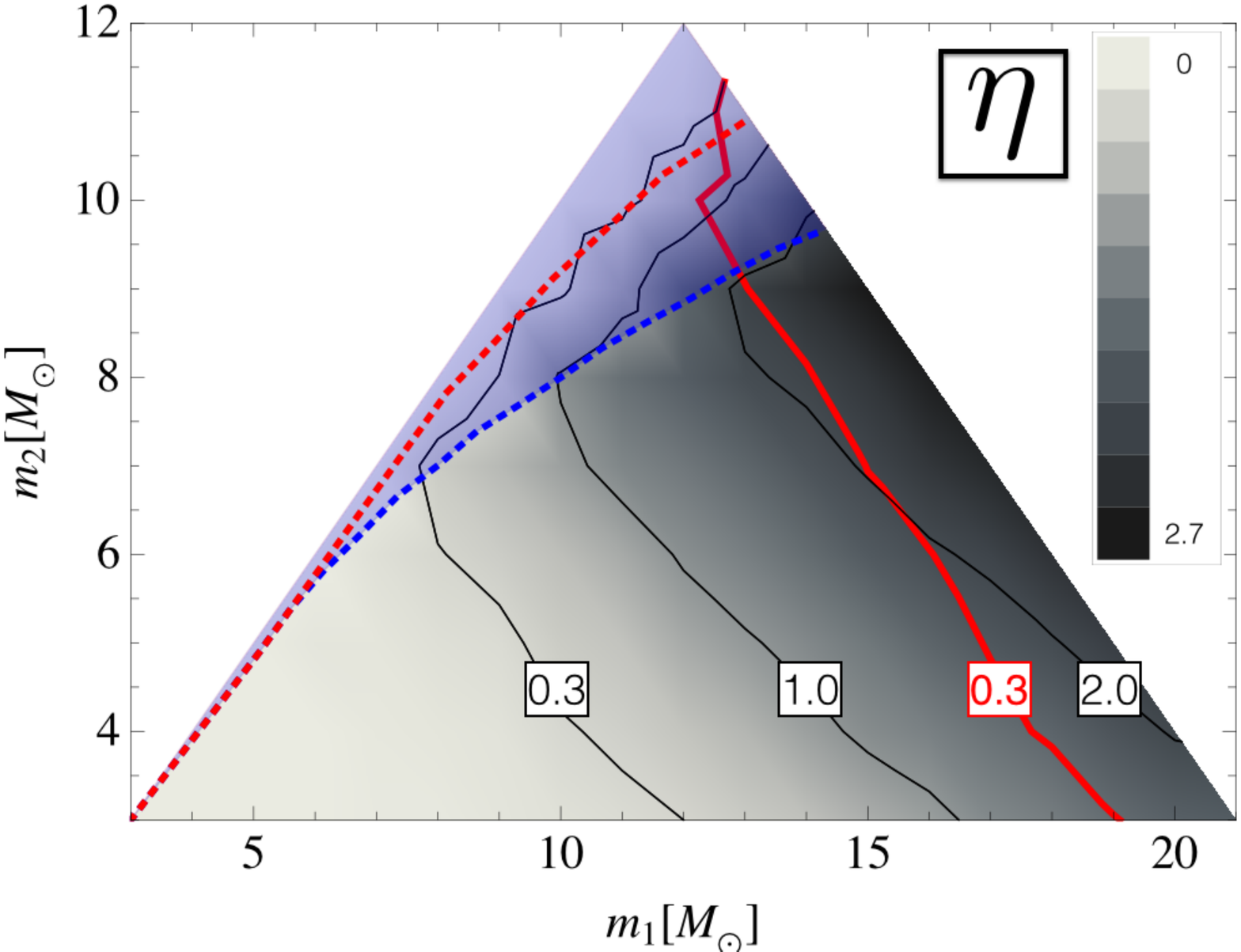}
\caption{\label{fig.frac-bias-isco}Fractional biases ($b/\sigma$) for $\Iisco$ templates  
for nonspinning BBHs, where $\sigma$ is the {\it true} statistical uncertainty given in figure~\ref{fig.error}.  
The blue dotted line denotes $\eta_{\rm crit}$.
For comparison, we include $\eta_{\rm crit}$ (red dotted line) and a fractional bias contour (red line) for $\Imerg$ templates taken from Figure~\ref{fig.frac-bias-merg}.}
\end{center}
\end{figure}

On the other hand, two results for $b_{\eta}/\eta$ were given in figure~5 of Ref.~\cite{Bos10} in low mass region, where one was obtained numerically from the overlap surfaces,
and the other was calculated by the analytic approximation described in \cite{Cut07}.
We find that the two results are quite different, the authors did not give sufficient explanations on  this apparent disagreement.
However, our result for $b_{\eta}/\eta$ can successfully explain the features in both results. 
The contours below $\eta_{\rm crit}$  show a pattern similar to that in the analytical result\footnote{
If we allow $\eta^{\rm rec}$ to range over the unphysical values, the contours 
will be smoothly extended to the blue shaded region, giving a pattern similar to that in the analytical result in the entire low mass region.}.
Above $\eta_{\rm crit}$, the biases are quite small near $\eta_0=0.25$ and do not appear to change even as $M$
is increased, this trend is consistent with that in the numerical result.

The fractional biases are summarised in figure~\ref{fig.frac-bias-isco},
and these are also several times larger than those for $\fmerg$ templates.
The valid criterion of the template bank to satisfy $b/\sigma < 1$ (where SNR=20)
is obtained in the range of $M<17 \msun$.
Therefore, the critical mass for $\Iisco$ templates for the parameter estimation efficiency
is determined by $\Mcrit \sim  17 \msun$.

%=================  SNR dependence	====================

\subsection{Statistical uncertainties for inspiral templates and the SNR dependence} \label{sec.3.4}

\begin{figure}[t]
\begin{center}
\includegraphics[width=5cm]{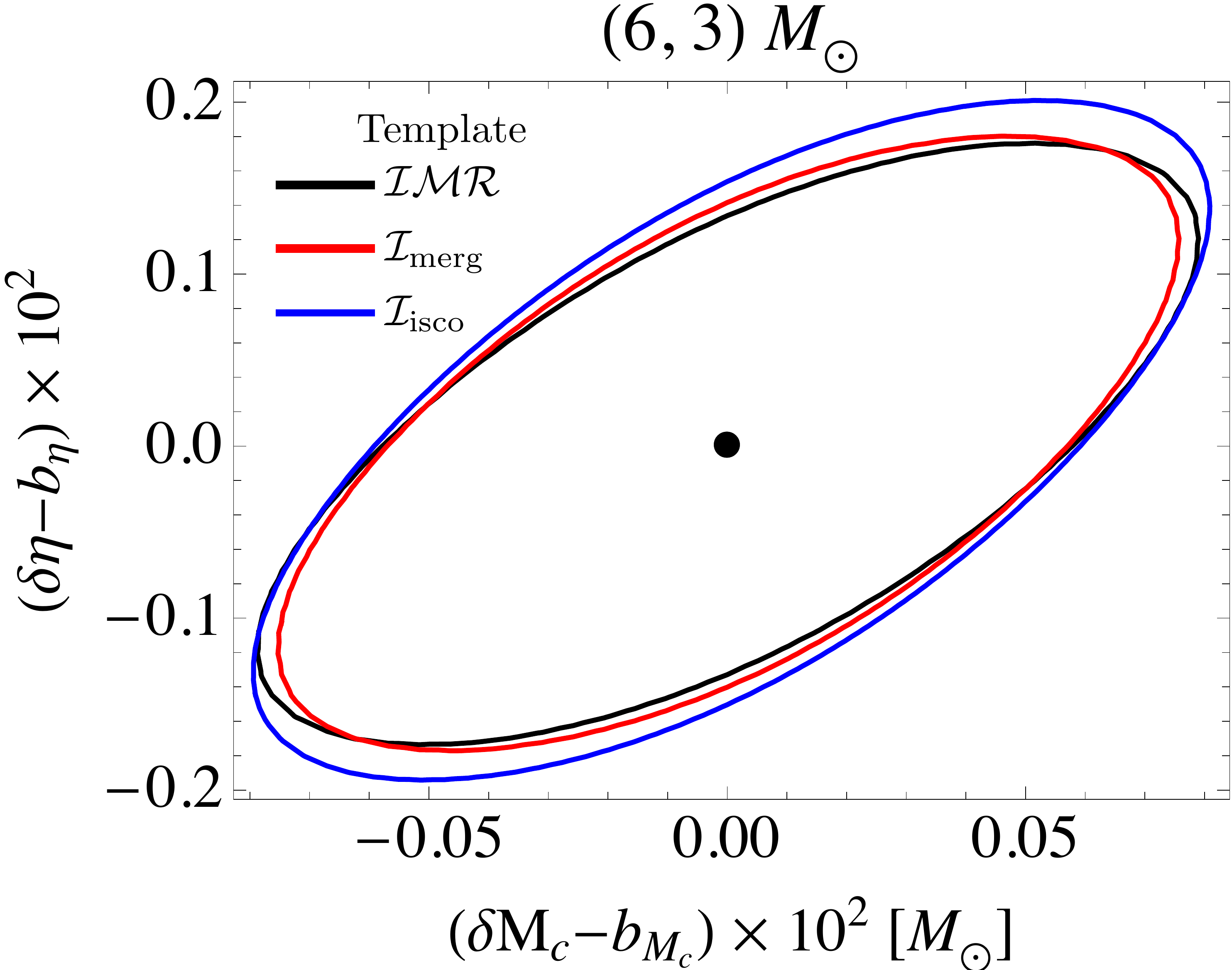}
\includegraphics[width=5cm]{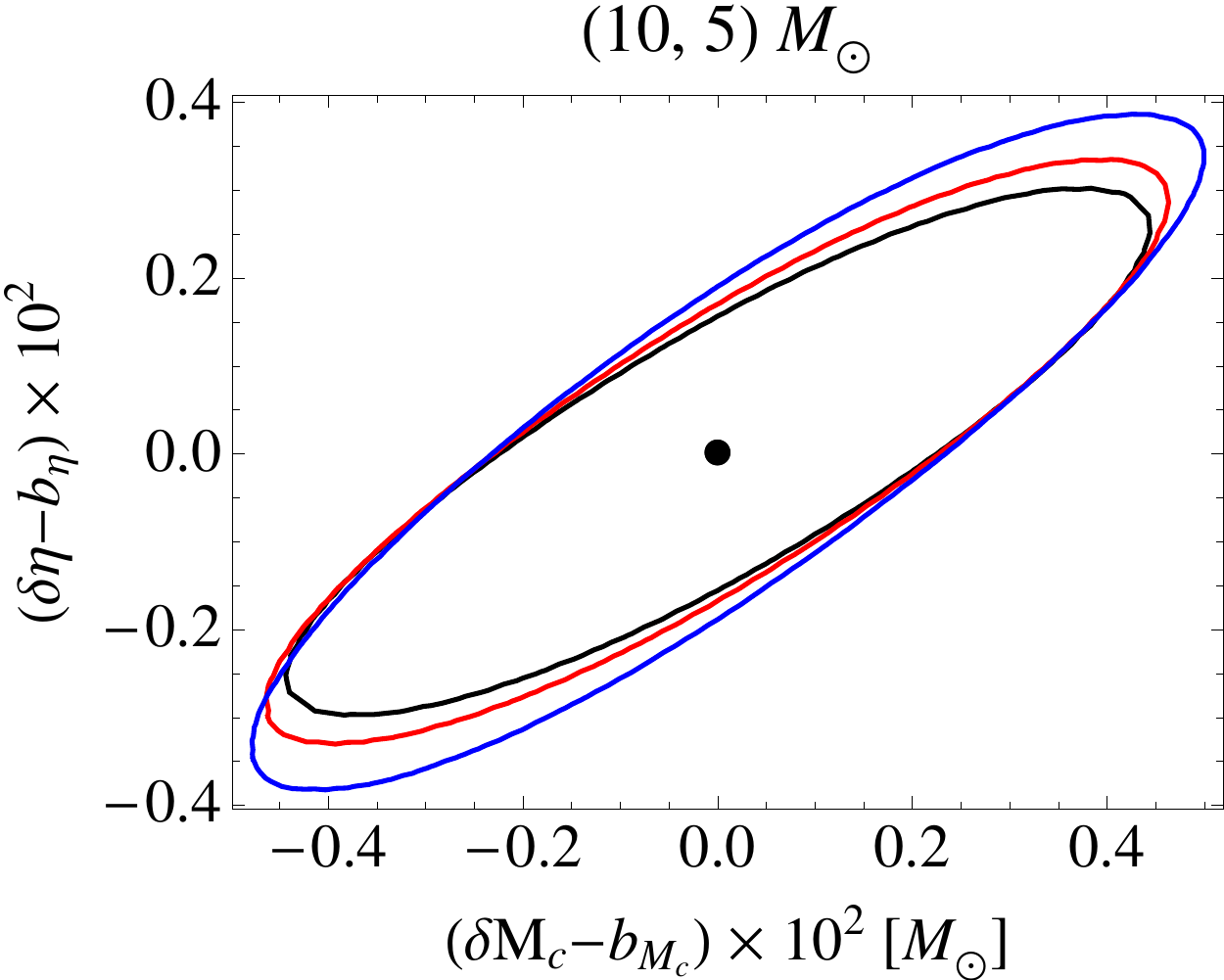}
\includegraphics[width=5cm]{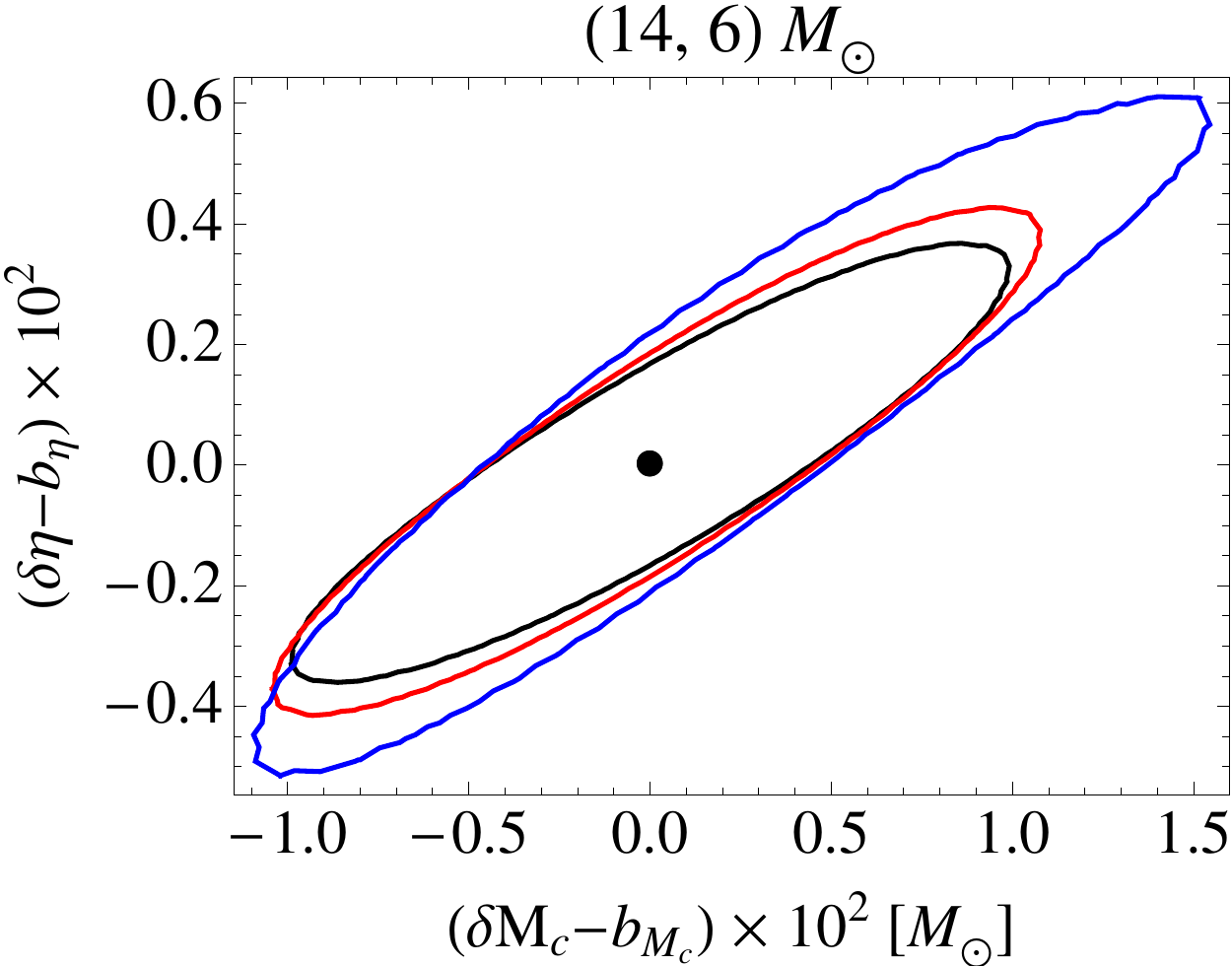}
\caption{\label{fig.biased-overlap-comparison}Comparison between overlaps for three template models. 
All contours correspond to $\hat{P}=0.99714$ (SNR=20), where $\hat{P}$ is an overlap weighted by FF.
}
\end{center}
\end{figure}

When using inaccurate template waveforms, not only the bias is produced but also the statistical uncertainty possibly differs from the {\it true} statistical uncertainty.
To see this, we choose three binaries with masses (6, 3), (10, 5) and (14, 6) $\msun$, and
depict their overlap contours  for $\IMR$ (black), $\Imerg$ (red) and $\Iisco$ (blue) templates together in figure~\ref{fig.biased-overlap-comparison},
where we define the axes by $\delta \lambda -b_{\lambda}$ so that the three overlaps
arrange at (0, 0).
The  contours correspond to $\hat{P}=0.99714$, and following equation (\ref{eq.P}) this corresponds to the two-dimensional confidence region with SNR=20, so they indicate the same level of confidence region.
Here, we define $\hat{P}$ as an overlap weighted by FF,  
\be
\hat{P} \equiv {P \over {\rm FF}},
\ee
so that the biased overlap surfaces for $\Imerg$ and $\Iisco$ templates also have a maximum value of $1$.
We see that the  red contours are slightly larger than the black contours,
and this difference  tends to increase as the binary mass increases. 
However, the  red contours are  nearly comparable in size with the black contours for all models.
This implies that the biased statistical uncertainties ($\sigma^{\rm biased}$) are comparable to the {\it true} statistical uncertainties ($\sigma$)
in our low mass region, but this agreement is not guaranteed at very high mass region.
On the other hand, the blue and black contours are comparable in size for a low mass binary,
but  the  difference between the two rapidly increases with increasing $M$. 
Therefore, the biased statistical uncertainties for $\Iisco$ can be reliable only if the binary masses are sufficiently low.

\begin{figure}[t]
\begin{center}
\includegraphics[width=7cm]{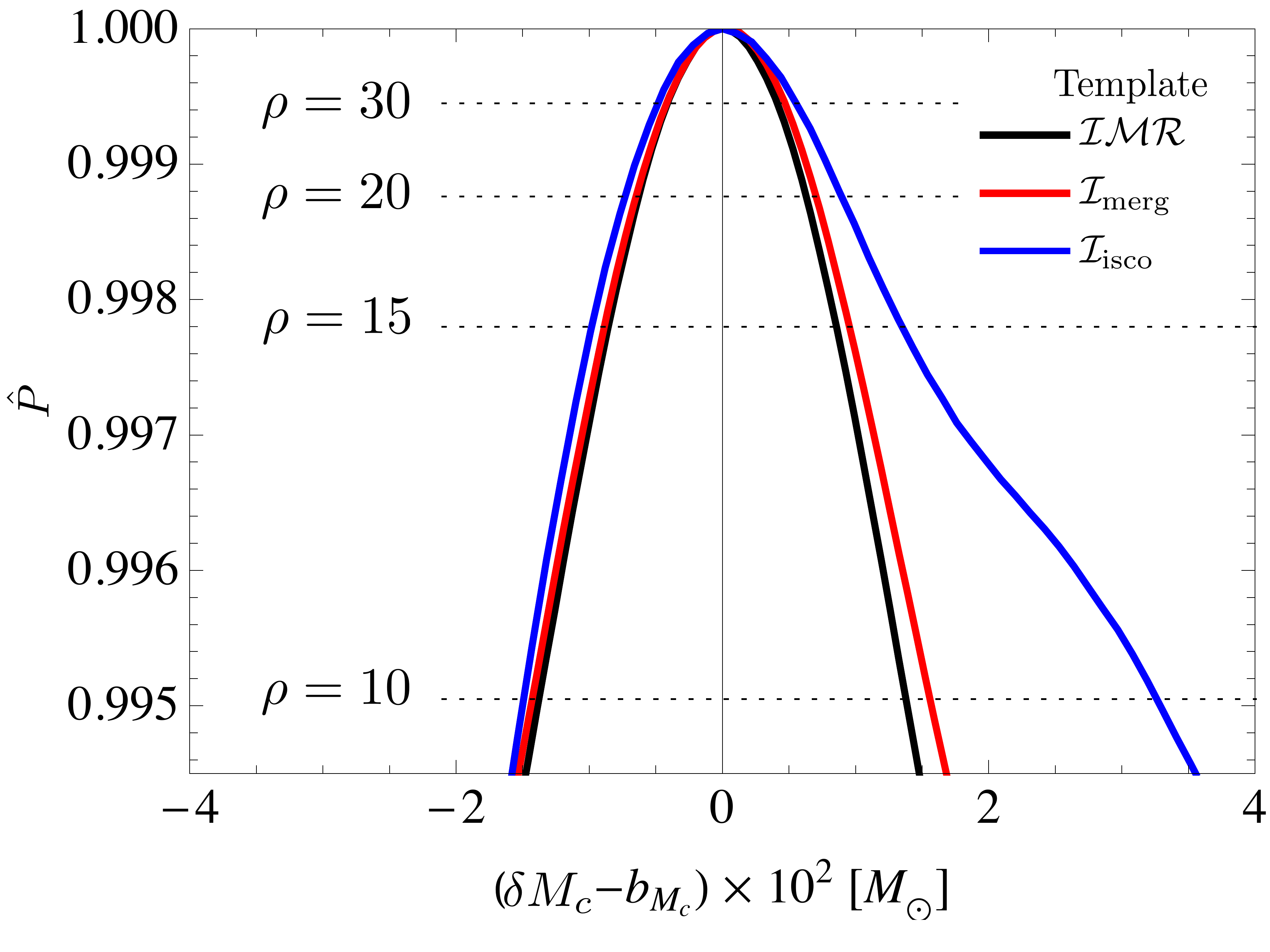}
\includegraphics[width=7cm]{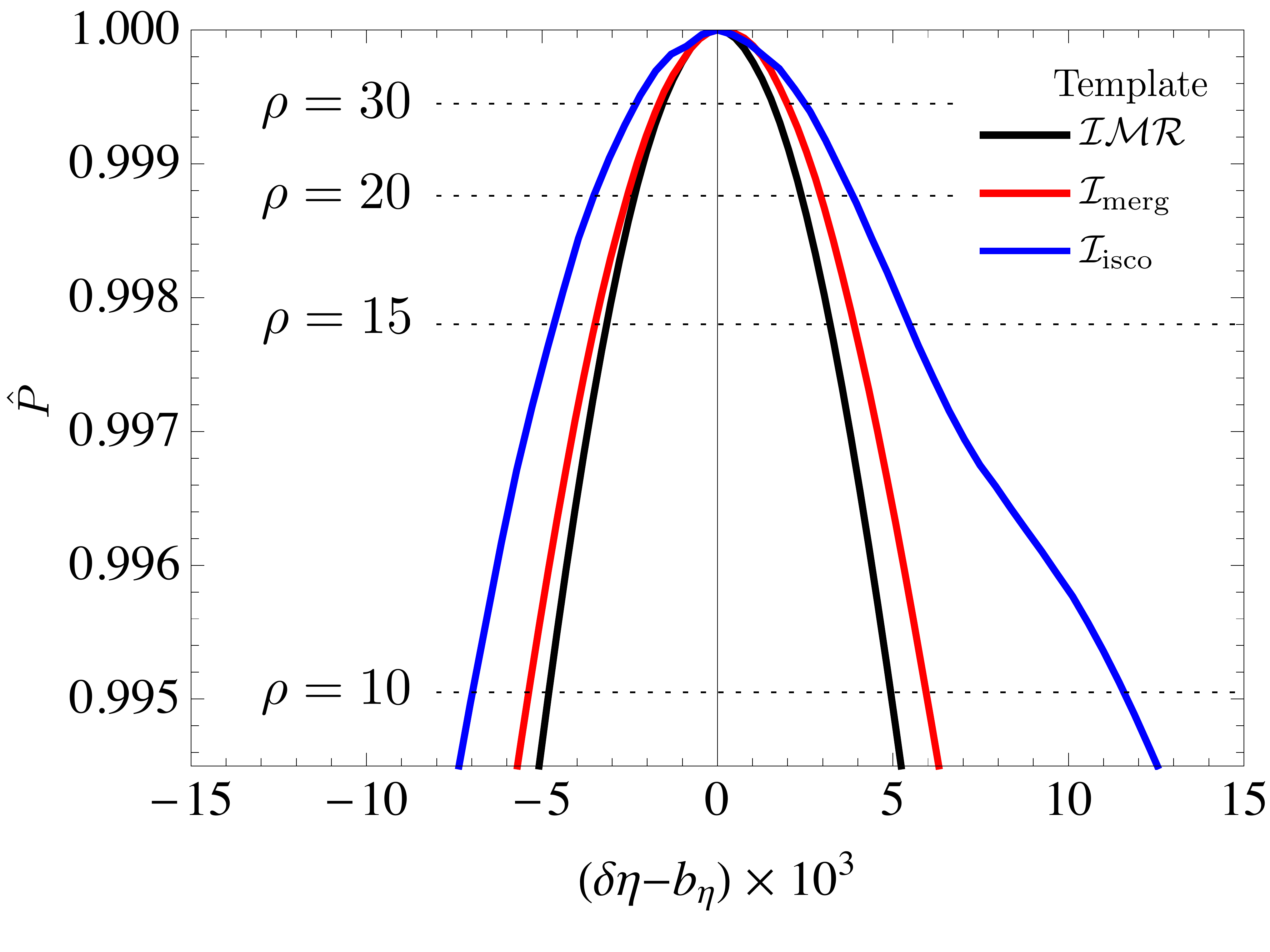}
\caption{\label{fig.CI-comparison}One-dimensional overlap distributions for the three template models. 
Dotted lines indicate $\hat{P}=0.99945, 0.99876, 0.99780$ and 0.99505, respectively.
These lines correspond to the lower boundaries of the overlap curves for calculation of the confidence intervals with the SNRs of 30, 20, 15 and 10, respectively (see figure~\ref{fig.CI} for more details). 
We use a binary with masses $(14, 6) \msun$.
Note that for $\Iisco$ templates, the blue curves are 
much wider than the black curves,
and the quadraticity is significantly broken at ${\rm SNR}=10$.
}
\end{center}
\end{figure}

So far, we have assumed a fixed SNR of $20$.
In order to investigate the dependence of our results on the SNR, we choose one source binary with masses $(14, 6) \msun$.
In figure~\ref{fig.CI-comparison}, we illustrate one-dimensional overlap distributions calculated by using $\IMR$, $\Imerg$ and $\Iisco$ templates, 
where we also define the $x$-axis by $\delta \lambda -b_{\lambda}$ so that the three overlaps
arrange at $x=0$.
The {\it true} uncertainty is determined by the black curve, and the
biased uncertainties are determined by the red and blue curves for $\Imerg$ and $\Iisco$ templates, respectively.
For a given SNR, the corresponding lower boundary of the overlap curves is denoted by the dotted line (see figure~\ref{fig.CI} for more details).
In the above, we showed that for a given SNR, the biased uncertainties ($\sigma^{\rm biased}$) obtained by the inspiral templates
can be considerably larger than the  {\it true} uncertainties ($\sigma$) if the binary masses are sufficiently high, and this behavior  
is much more significant for $\Iisco$ templates.
Similarly, figure~\ref{fig.CI-comparison} shows that if the SNR is too low,
$\sigma^{\rm biased}$ can be considerably larger than $\sigma$ even in low mass region,
especially for $\Iisco$ templates.
One can see that the difference between the black and red curves is overall small
for both $M_c$ and $\eta$.
However, the blue curves are 
much wider than the black ones at ${\rm SNR}=10$,
and the quadraticity is also broken.
We found that this discrepancy can be more significant for more massive binaries.
We therefore conclude that for $\Imerg$ templates $\sigma^{\rm biased}$ are overall acceptable
in our mass region for any SNRs above $\sim10$.
For $\Iisco$ templates, however, $\sigma^{\rm biased}$ can be acceptable only if the binary mass is sufficiently low and the SNR is sufficiently high.

\begin{figure}[t]
\begin{center}
\includegraphics[width=14cm]{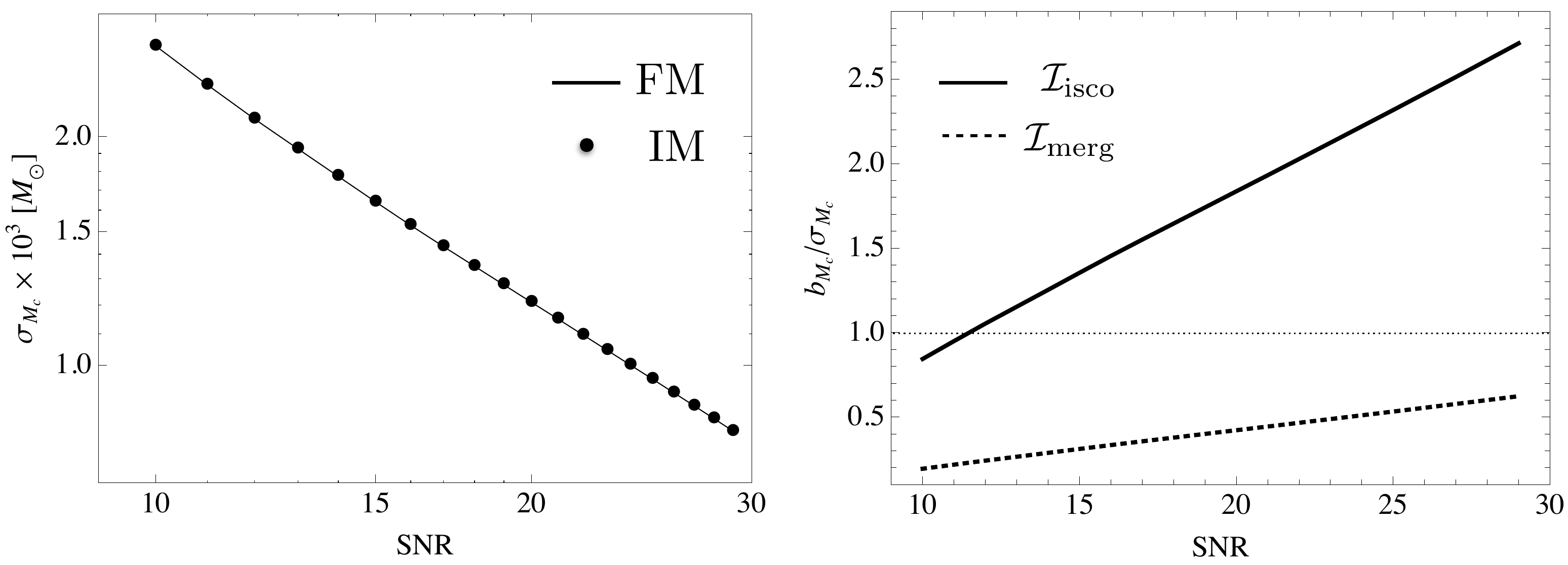}
\caption{\label{fig.snr-dependence} Comparison between parameter estimation uncertainties ($\sigma$) computed by the analytic FM and the IM methods with various SNRs (left) and the corresponding fractional biases ($b/\sigma$) for a bias calculated by an IMR signal and inspiral templates (right). We assume a binary with masses $(16, 8) \msun$ for the signal. 
}
\end{center}
\end{figure}

The FM formalism implies that the {\it true} statistical uncertainty ($\sigma$) is inversely proportional to the SNR. 
To see this from our overlaps, we calculate the confidence intervals using the IM method varying the SNR for a binary with masses $(16, 8) \msun$.
In the left panel of figure~\ref{fig.snr-dependence}, we compare the result with the FM result, and find a very good agreement between the two methods.
On the other hand, FF and systematic bias ($b_{\lambda}$) are independent of the SNR but
depend only on the template model ($h_t$).
Thus, for a bias calculated by an IMR signal and inspiral templates, the fractional bias ($b/\sigma$) increases with increasing SNR,
and those are illustrated in the right panel of figure~\ref{fig.snr-dependence} for $\Iisco$ and $\Imerg$ templates.
This result indicates that $\Imerg$ template model becomes much more efficient than $\Iisco$ for parameter estimation as the SNR increases.
For $\Iisco$ templates, if the SNR is lower than $\sim 12$, $b_{M_c}$ can be smaller than $\sigma_{M_c}$,
but the statistical uncertainty will not be acceptable because
the quadraticity of the biased overlap can be significantly broken at such a low SNR as seen in figure~\ref{fig.CI-comparison}.

%=================    spin 	====================
\subsection{Aligned-spinning case}

\begin{figure}[t]
\begin{center}
\includegraphics[width=\columnwidth]{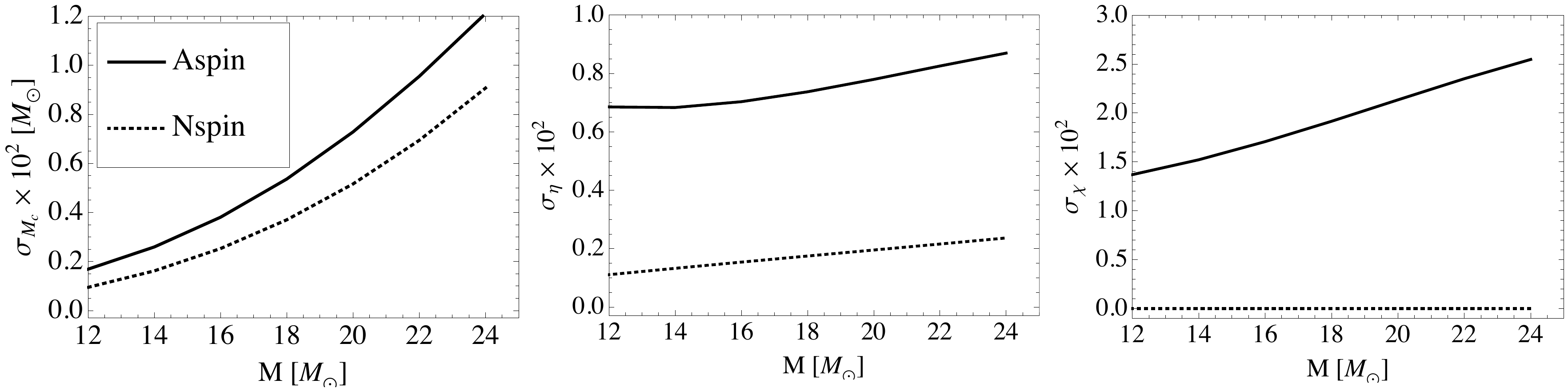}
\caption{\label{fig.error-comparison}  Statistical uncertainties for nonspinning (Nspin) and aligned-spin (Aspin) binaries calculated by using the FM method.
We use PhenomA  for non spinning and PhenomC for aligned-spin models assuming fixed mass ratio $(m_1/m_2=3)$ and spin ($\chi=0.5$).}
\end{center}
\end{figure}

\begin{figure}[t]
\begin{center}
\includegraphics[width=10cm]{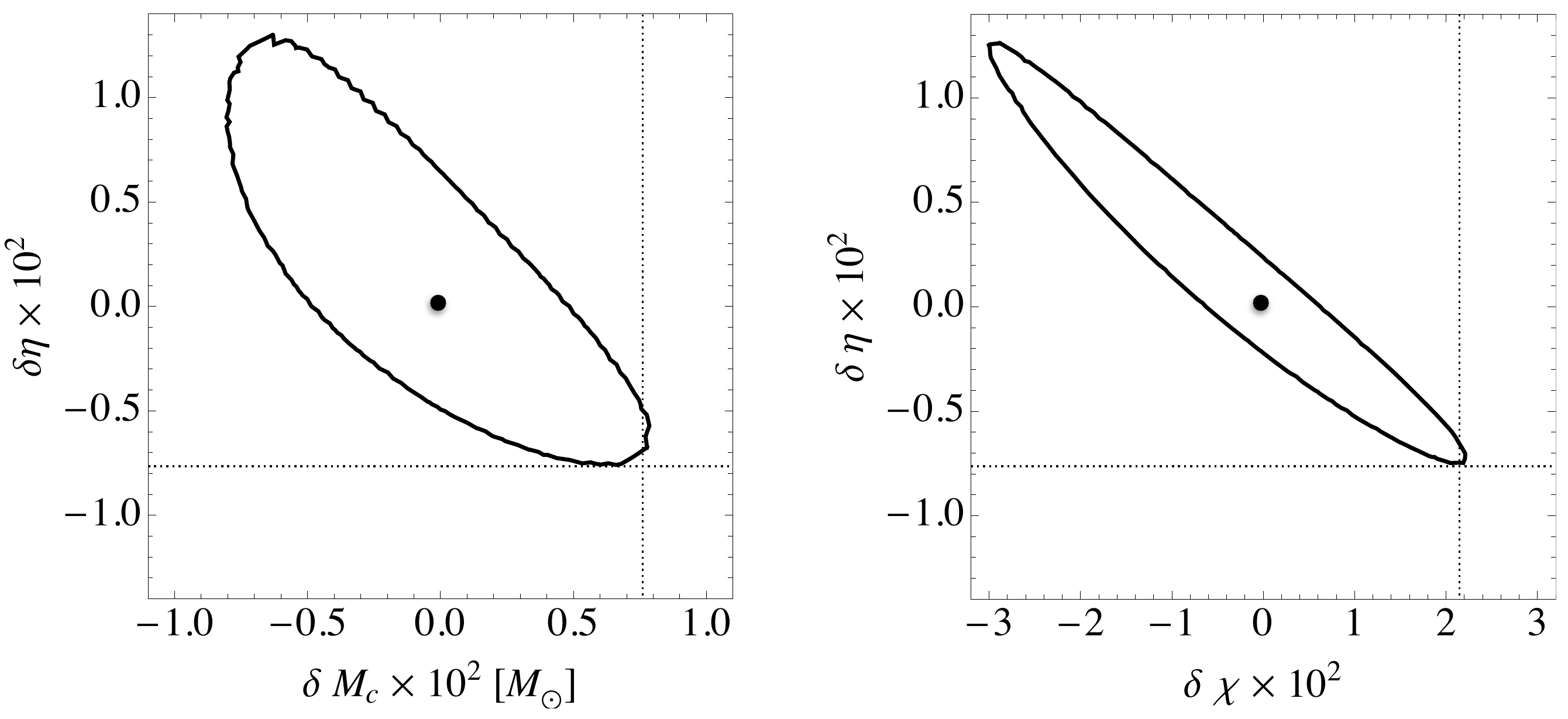}
\caption{\label{fig.3doverlap}  Comparison of statistical uncertainties between IM and FM methods  for a aligned-spin binary with $(15, 5)~\msun$ and $\chi=0.5$ using PhenomC model. The contours indicate $P=0.99876$, and the dotted lines indicate the uncertainties obtained from the FM.}
\end{center}
\end{figure}

In this subsection, we show some results for aligned-spinning BBHs.
We choose to use PhenomC~\cite{San10} because that is  the most recent model for this system.
In appendix, we briefly describe this model.
In order to include the spin effect, PhenomC has the effective spin parameter $\chi\equiv (1+\delta)\chi_1/2+(1-\delta)\chi_2/2$ where $\delta \equiv (m_1-m_2)/M$ and $\chi_i \equiv S_i/m_i^2$, $S_i$ being the spin angular momentum of the {\it i}th BH.
In this work, we  consider only one  value ($\chi=0.5$) for the spin parameter,
and vary total masses from $12\msun$ to $24\msun$ for a fixed mass ratio of $m_1/m_2=3$.
First, we calculate statistical uncertainties for mass and spin parameters.
Since we have seen that FM is sufficiently accurate
in estimating the statistical uncertainties for PhenomA,
we also adopt FM method for PhenomC. 
In figure~\ref{fig.error-comparison}, we compare the uncertainties ($\sigma$) for aligned-spinning binaries with those given in figure~\ref{fig.error} for nonspinning binaries.
We find that $\sigma_{M_c}$ for both systems rapidly increase with increasing $M$, but
$\sigma_{M_c}$ for aligned-spinning BBHs are a bit larger than those for nonspinning BBHs within a factor of 2.
Uncertainties in $\eta$ also slowly increase for both systems, but the difference in $\sigma_{\eta}$ between the two systems 
is much more significant compared to the case for $\sigma_{M_c}$.
This is due to the degeneracy between the symmetric mass ratio and the effective spin~\cite{Bai13},
which is already well known in PN theory~\cite{Cut94,Poi95}.
Thus, if we project the three-dimensional confidence region onto the ($\eta-\chi$) plane,
we can see a long ellipse, that has a strong correlation  between the two parameters (see, figure~\ref{fig.3doverlap}). 
For reference, we also present $\sigma_{\chi}$, where we find that the uncertainties almost linearly increase.

For a sanity check between the IM and the FM methods for the aligned-spinning system, we compute the three-dimensional overlap ellipsoid for the binary of $(15, 5) ~\msun$ with $\chi=0.5$,
and calculate the statistical uncertainties for the mass and spin parameters by using the IM method.
In figure~\ref{fig.3doverlap}, we illustrate the overlap contours of $P=0.99876$ in the ($M_c - \eta$) and ($\chi - \eta$) planes, respectively.
The doted lines indicate the statistical uncertainties obtained from the FM method, i.e., $\{\sigma_{M_c}, \sigma_{\eta}, \sigma_{\chi}\}\simeq\{0.0073, 0.0078, 0.0213\}$.
We find that the uncertainties for both methods are in good agreement for all parameters if we consider the one-sided overlap.
However, while the two-dimensional overlaps calculated by using PhenomA model are nearly quadratic for all SNRs above 10 as described in figure~\ref{fig.snr-dependence},
the three-dimensional overlap are less quadratic even at SNR=20. In addition, this behavior can be more pronounced as the SNR decrease (e.g., see figure~8 of Ref.~\cite{Pur13}).
Therefore, for the aligned-spinning system, one should be careful about the choice of SNR in order to apply the IM and FM methods to the parameter estimation. This will be studied in detail in a future work.

\begin{figure}[t]
\begin{center}
\includegraphics[width=\columnwidth]{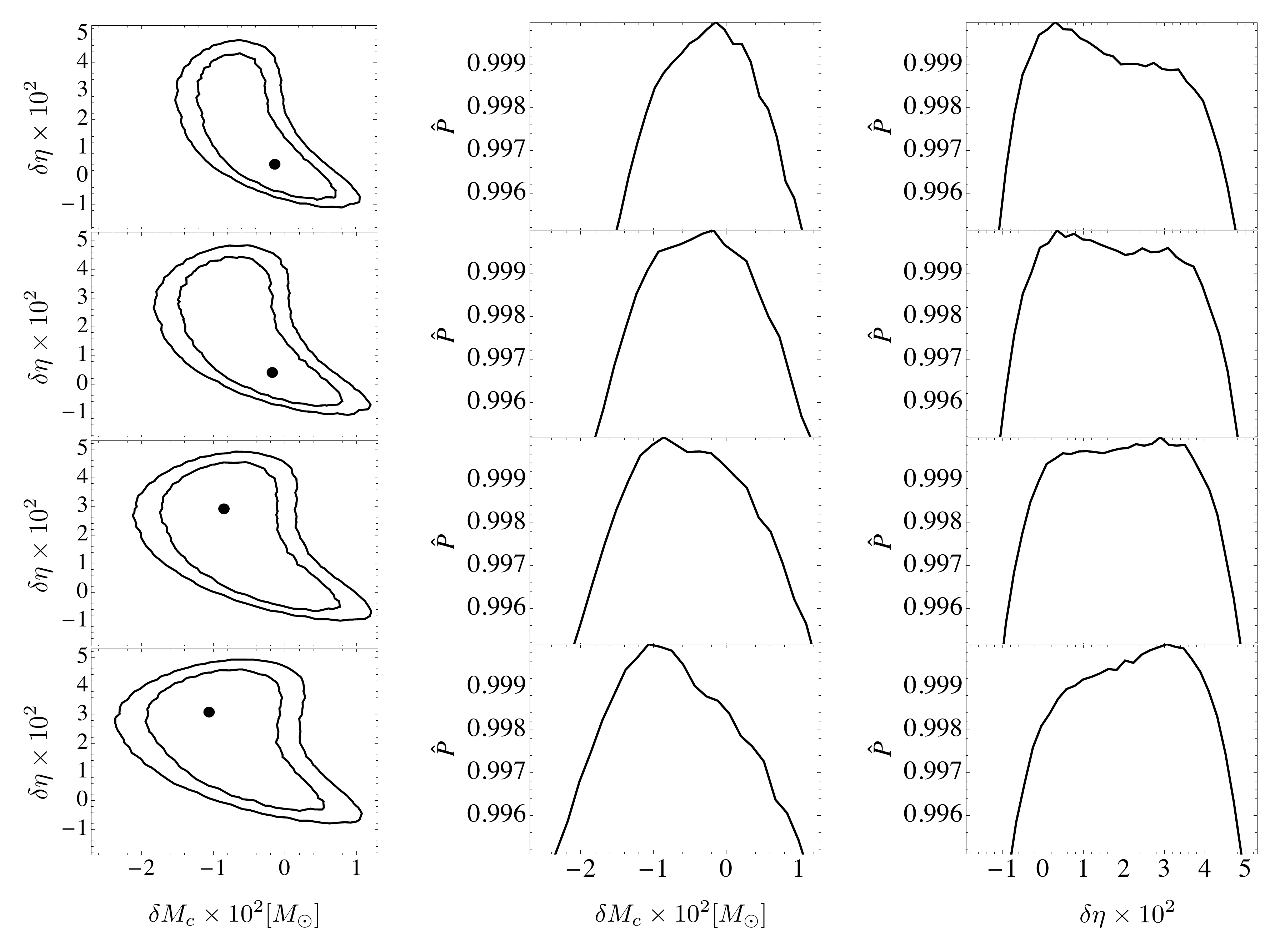}
\caption{\label{fig.3d-comparison} Confidence regions at SNR=15 and 20 calculated by using inspiral-PhenomC templates terminated at $\fmerg$ (left), and their projections onto $M_c$ (middle) and $\eta$ (right) axes. Total masses are 18, 19, 20 and 21$\msun$ from top to down.  Mass ratio and spin are fixed to be $m1/m2=3$ and $\chi=0.5$. Large dots indicate the recovered parameters ($M_c^{\rm rec},  \eta^{\rm rec}$).}
\end{center}
\end{figure}

Next, we calculate three-dimensional overlap distributions with IMR PhenomC signals and inspiral PhenomC templates.
In PhenomC, since the amplitude of inspiral-merger phase is defined by one smooth function  
without a transition frequency, we terminate the PhenomC waveforms at $\fmerg$ defined in equation~(\ref{eq.phenomparameters}) to obtain the inspiral-PhenomC template waveforms.
We found that all of the three-dimensional confidence regions had the long-thin-curved banana shapes in the ($M_c, \eta, \chi$) space.
We give some examples in figure~\ref{fig.3d-comparison}, where we show the two-dimensional confidence regions by projecting the original three-dimensional overlaps
onto the ($M_c, \eta$) plane, and the one-dimensional overlap functions  by projecting the two-dimensional confidence regions
onto each axis.
From these overlap distributions, we find an interesting behavior in the variation of biases.
The recovered parameter ($M_c^{\rm rec},  \eta^{\rm rec}$) is located at the bottom right-hand side of the contours at $18\msun$, and that is slightly moved at $19\msun$.
However, that is suddenly moved to the top left-hand side at $20 \msun$, and again slightly moved at $21\msun$.
This sudden movement is well described in the one-dimensional overlap functions, where
$M_c^{\rm rec}$  tends to move from right to left, and  $\eta^{\rm rec}$ moves oppositely.
It seems that the overlap surface becomes bimodal at masses between $19 \msun$ and $20 \msun$,
and then the position of maximum overlap suddenly moves from one peak to the other peak.

\begin{figure}[t]
\begin{center}
\includegraphics[width=14cm]{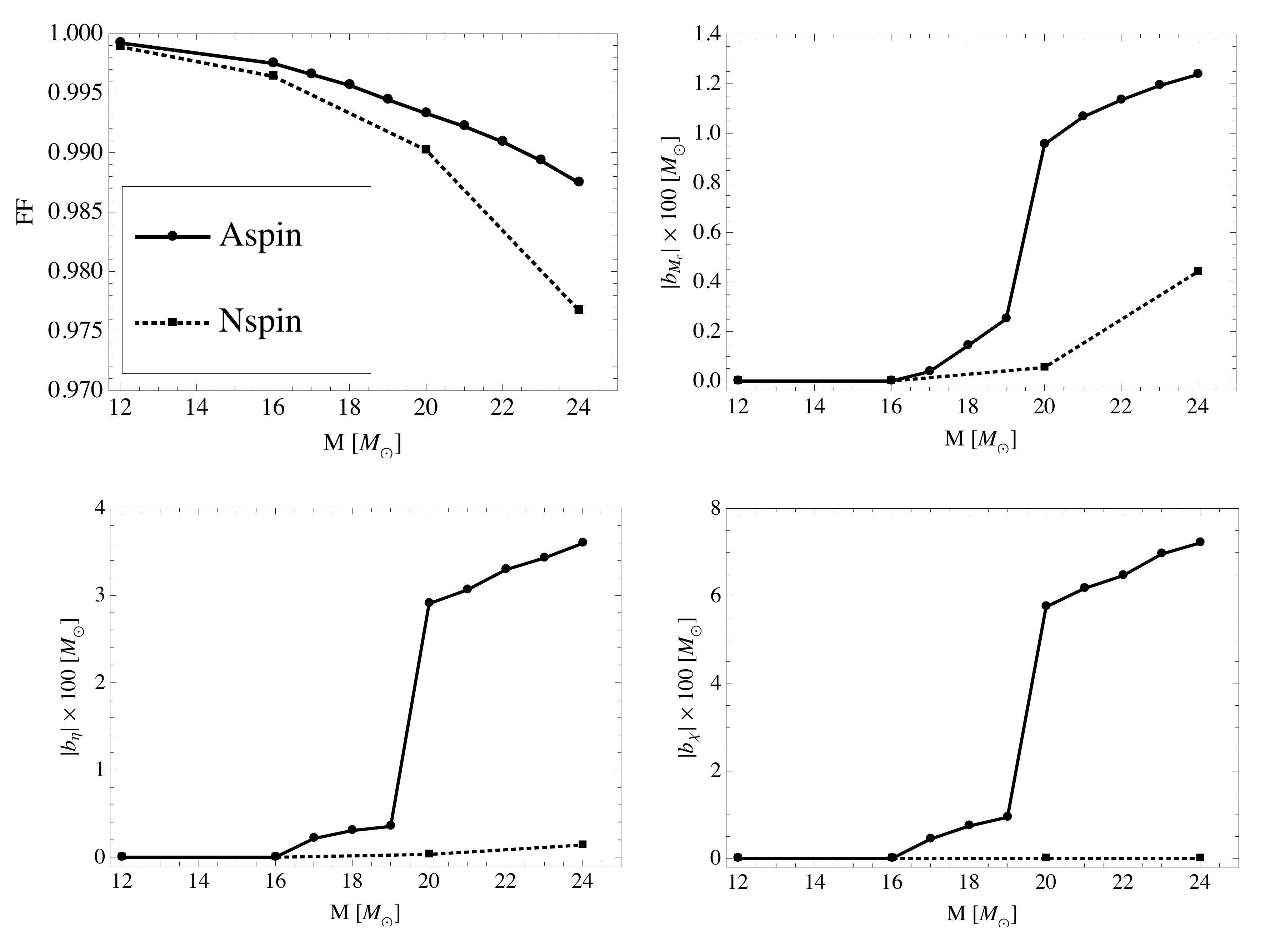}
\caption{\label{fig.bias-comparison} Fitting factors and biases for nonspinning (Nspin) and aligned-spin (Aspin) binaries  calculated by varying total masses with fixed mass ratio $(m_1/m_2=3)$ and spin ($\chi=0.5$).  Note a sudden jump up of biases between $M=19 \msun$ and $20 \msun$ (see figure~\ref{fig.3d-comparison}).}
\end{center}
\end{figure}

By exploring the {\it complicated} three-dimensional overlap spaces, we investigate FFs and biases for the inspiral PhenomC templates\footnote{The procedure to find the overlap distribution $\hat{P}>0.995$ is the same as in the case for the nonspinning system, but we use $31 \times 31 \times 31$ grid points in the final grid search.}.
In figure~\ref{fig.bias-comparison}, we compare the results with those  given in figures~\ref{fig.ff-merg} and \ref{fig.bias-merg} for the nonspinning inspiral templates ($\Imerg$).
In the top left panel, we find that FFs for the  aligned-spinning system are better than those of the nonspinning system for all masses considered here.
This improvement is just due to the expansion of the parameter space from two to three dimensions.
Since the masses are strongly correlated with the spin, the addition of the spin dimension to the overlap space
can increase  FF compared to the two-dimensional searches.
The other three panels show biases for nonspinning and aligned-spinning systems.
We find that biases are also much larger for the aligned-spinning system.
In particular, we see that while  the biases increase gradually with increasing $M$ for the nonspinning system,
those for the aligned-spinnig system suddenly jump up at between $19 \msun$ and $20 \msun$
as described above.

%=================   section 4 : summary 	====================
\section{Summary and future work}\label{sec4}

Making use of phenomenological waveform model (PhenomA),
we defined the IMR model $\IMR$ and the two inspiral models $\Imerg$ and $\Iisco$,
and assumed $\IMR$ as our complete signal model.
We described how to calculate the statistical uncertainties in parameter estimation from the overlap surfaces
and calculated the uncertainties using $\IMR$ templates for nonspinning BBHs in the range of $M \leq 30 \msun$ with Advanced LIGO sensitivity.
We investigated the validity of the inspiral 
templates in detection and parameter estimation, respectively, and provided various crucial values in detail.
The results of this work can be summarised as

\begin{itemize}
\item  For $\IMR$ templates, statistical uncertainties ($\sigma_{\lambda} / \lambda$)
overall depend on the chirp mass of the system and weakly depend on 
the mass ratio in highly asymmetric mass region.
The percentage uncertainties for $M_c$ and  $\eta$ are  
$0.0067 \% - 0.22 \%$ and  $0.43 \% - 2.0 \%$, respectively with a SNR of 20,
and these results are in good agreement with the FM estimates of uncertainty within $\sim 2\%$ differences.

\item For $\Imerg$ templates,
the valid criterion of the template bank to satisfy FF $\geq0.97$
is obtained in the range of $M < 24 \msun$.
The bias increases with increasing mass more rapidly than the statistical uncertainty,
that begins to exceed the uncertainty at $M \sim 26 \msun$.
Thus, $\Mcrit \sim 24 \msun$ and $\sim 26 \msun$ for the detection and the parameter estimation, respectively.

\item For $\Iisco$ templates,
$\Mcrit \sim 15 \msun$ and $\sim 17 \msun$ for the detection and the parameter estimation, respectively.

\item In our mass region, the biased statistical uncertainties ($\sigma^{\rm biased}$) calculated by the $\Imerg$ templates are slightly larger than the {\it true} uncertainties ($\sigma$) but overall comparable. However, those for $\Iisco$ are acceptable only if the binary masses are sufficiently small.
The difference between $\sigma$ and $\sigma^{\rm biased}$ tends to increase with increasing mass or decreasing SNR.

\end{itemize}

We also showed some results for aligned-spinning binaries with fixed mass ratio ($m_1/m_2=3$) and spin ($\chi=0.5$). 
We used the waveform model PhenomC, and  calculated statistical uncertainties ($\sigma_{\lambda}$) using the FM method.
Aligned-spinning inspiral waveforms were obtained by terminating the PhenomC waveforms at $\fmerg$. 
For these binaries we found that

\begin{itemize}
\item  For IMR PhenomC templates, $\sigma_{M_c}$ for aligned-spinning binaries are a bit larger than those for nonspinning binaries within a factor of 2.
However, the differences in $\sigma_{\eta}$ between the two systems are found to be much larger compared to the case for $\sigma_{M_c}$ due to the mass ratio-spin degeneracy.  

\item For inspiral-PhenomC templates, three-dimensional confidence regions have long-thin-curved banana shapes in the  ($M_c, \eta, \chi$) space.
FFs for the aligned-spinning system can be better than those for the nonspinning system, but biases are  much larger.
In particular,  the confidence regions can have bimodal distributions for the binaries with masses between $19\msun$ and $20\msun$.

\end{itemize}

In this work, we considered limited binary models for the aligned-spinning system,
so our results may not be generalised to those binaries with generic masses and spins.
We will extend our approach to generic aligned-spinning binaries  in a future work.
We showed that the analytic FM method is reliable for estimating the statistical uncertainties for the nonspinning system
because the  overlap surface is nearly quadratic in the ($M_c, \eta$) plane,
However, the three-dimensional overlap was found to be less quadratic even at a high SNR of 20 for our binary model.
We will also investigate the validity of the FM method for generic aligned-spinning BBHs in detail.
The mass-spin degeneracy in the aligned-spinning system generally limits our ability to measure the individual component masses~\cite{Han13}, but
the degeneracy can be broken in the precessing binaries~\cite{Cha15}.
This can also be studied by comparing the FM results for both binary systems.

%%%%%%%%%%%%%%%%%%%%%%%%%%%%%%%%%%%%%%%%%%%%%%%%%%%%%%%%%%%

%=======	Acknowledgements ===========================	
%

\ack{This work used the computing resources at the KISTI Global Science Experimental Data Hub Center (GSDC).}

%%%%%%%%%%%%%%%%%%%%%%%%% appendix %%%%%%
\appendix
\section*{Appendix}
\setcounter{section}{1}
The wave amplitude of PhenomC terminates at $\fcut=0.15/M$, and that  is constructed from  two parts as
\be
A_{\rm eff}=A_{\rm PM}(f)w^-_{f_0}+A_{\rm RD}(f)w^+_{f_0},
\ee
where $A_{\rm PM}$ is the premerger amplitude calculated by a PN inspiral amplitude with the addition of a higher order frequency term:
\bea
A_{\rm PM}(f)=A_{\rm PN}(f)+\gamma_1f^{5/6}, \\
A_{\rm PN}=C\Omega^{-7/6}\left(1+\sum^5_{k=2}\gamma_k\Omega^{k/3} \right),
\eea
where $\Omega=\pi M f$, and
$A_{\rm RD}$ is the ringdown amplitude:
\be
A_{\rm RD}=\delta_i {\cal L'}[f,f_{\rm RD}(a,M),\delta_2 Q(a)]\bar{\sigma})f^{-7/6},
\ee
where the Lorentzian function is defined by ${\cal L'}(f,f_0,\bar{\sigma})\equiv \bar{\sigma}^2/[(f-f_0)^2+\bar{\sigma}/4]$,
and $Q$ is the quality factor which depends on the final BH spin $a$.
The two amplitude parts can be combined by tanh-window functions:
\be
w^{\pm}_{f_0}={1 \over 2}\left[ 1\pm{\rm tanh} \left({4(f-f_0) \over d} \right) \right],
\ee
where $d=0.005$. The transition frequency $f_0$ is determined by 
$f_0=0.98 f_{\rm RD}$ where $f_{\rm RD}$ is a ringdown frequency given in terms of $M$ and $a$.
The effective phase is calculated by a complete SPA inspiral phasing $\psi_{\rm SPA}$, a premerger phasing $\psi_{\rm PM}$ and a ringdown phasing $\psi_{\rm RD}$ as
\be
\Psi_{\rm eff}(f) = \psi_{\rm SPA}w^-_{f_1}+\psi_{\rm PM}w^+_{f_1}w^-_{f_2}+\psi_{\rm RD}w^+_{f_2},
\label{eq.effectivephase}
\ee
with $f_1=0.1f_{\rm RD}, f_2=f_{\rm RD}$ using $d=0.005$ in the window functions.
The premerger and ringdown phasing have the forms
\bea
\psi_{\rm PM}&=&{1\over \eta} (\alpha_1 f^{-5/3}+\alpha_2 f^{-1}+\alpha_3 f^{-1/3} +  \alpha_4+\alpha_5 f^{2/3}+\alpha_6 f),  \\
\psi_{\rm RD}&=&\beta_1+\beta_2 f,
\eea
where the $\alpha_k$ coefficients are inspired by the SPA phase, redefined and phenomenologically fitted to agree with the PN-NR hybrid waveforms,
while $\beta_{1,2}$ parameters are not fitted but obtained from the premerger ansatz by taking the value and slope of the phase at the transition point   $f_{\rm RD}$.
The coefficients  introduced in this model  are expressed in terms of $\eta$ and $\chi$, and those are
given in table 2 of~\cite{San10}.

%
%
%=======	Bibliography     ===========================	
%

\end{document}